\documentclass[prb,twocolumn,showpacs,amsmath,amssymb,superscriptaddress]{revtex4}

\usepackage{graphicx}
\usepackage{dcolumn}
\usepackage{bm}
\usepackage{color}

\usepackage{epsfig,float,afterpage,amssymb,wrapfig,psfrag}
\usepackage{subfigure}

\DeclareMathAlphabet{\bi}{OML}{cmm}{b}{it}
\newcommand{\half}{\text{$\textstyle\frac{1}{2}$}}
\newcommand{\hloc}{h_{\text{loc}}}
\newcommand{\kk}{\mathbf{k}}
\newcommand{\loc}{\mathrm{loc}}
\newcommand{\pdag}{\phantom{\dag}}
\newcommand{\ph}{$p$-$h$}
\newcommand{\stat}{\mathrm{stat}}

\newcommand{\chistat}{\chi_{\stat}}
\newcommand{\tot}{\text{tot}}

\begin{document}
\title{Quantum criticality in the pseudogap Bose-Fermi Anderson and Kondo
models: Interplay between fermion- and boson-induced Kondo destruction}

\author{J.\ H.\ Pixley}
\affiliation{Department of Physics and Astronomy, Rice University,
Houston, Texas, 77005, USA}
\author{Stefan Kirchner}
\affiliation{Max Planck Institute for the Physics of Complex Systems,
01187 Dresden, Germany}
\affiliation{Max Planck Institute for Chemical Physics of Solids,
01187 Dresden, Germany}
\author{Kevin Ingersent}
\affiliation{Department of Physics, University of Florida, Gainesville, Florida
32611-8440, USA}
\author{Qimiao Si}
\affiliation{Department of Physics and Astronomy, Rice University,
Houston, Texas, 77005, USA}

\date{\today}
\begin{abstract}
We address the phenomenon of critical Kondo destruction in pseudogap Bose-Fermi
Anderson and Kondo quantum impurity models. These models describe a localized
level coupled both to a fermionic bath having a density of states that vanishes
like $|\epsilon|^r$ at the Fermi energy ($\epsilon=0$) and, via one component
of the impurity spin, to a bosonic bath having a sub-Ohmic spectral density
proportional to $|\omega|^s$. Each bath is capable by itself of suppressing the
Kondo effect at a continuous quantum phase transition. We study the interplay
between these two mechanisms for Kondo destruction using continuous-time
quantum Monte Carlo for the pseudogap Bose-Fermi Anderson model with
$0<r<\half$ and $\half\le s<1$, and applying the numerical renormalization-group
to the corresponding Kondo model. At particle-hole symmetry, the models exhibit
a quantum critical point between a Kondo (fermionic strong-coupling) phase
and a localized (Kondo-destroyed) phase. The two solution methods, which
are in good agreement in their domain of overlap, provide access to the
many-body spectrum, as well as to correlation functions including, in
particular, the single-particle Green's function and the static and dynamical
local spin susceptibilities. The quantum-critical regime exhibits the
hyperscaling of critical exponents and $\omega/T$ scaling in the dynamics that
characterize an interacting critical point. The $(r,s)$ plane can be divided
into three regions: one each in which the calculated critical properties are
dominated by the bosonic bath alone or by the fermionic bath alone,
and between these two regions, a third in which the bosonic bath governs the
critical spin response but both baths influence the renormalization-group flow
near the quantum critical point.
\end{abstract}


\maketitle

\section{Introduction}

Quantum criticality is currently being pursued in a number of strongly
correlated materials ranging from intermetallic rare earth compounds to
insulating magnets and even engineered systems. This interest is spurred in
part by the observation that unconventional superconductivity commonly occurs
near quantum critical points, as well as by mounting experimental
evidence of the inadequacy of the traditional theoretical approach to
continuous transitions at absolute temperature $T=0$. This approach,
commonly referred to as the spin-density-wave or Hertz-Millis-Moryia
picture, is based on an extension to zero temperature of Landau's theory of
order-parameter fluctuations.\cite{Sachdev} The evidence for the breakdown of
the spin-density-wave picture is particularly compelling in the context of
intermetallic rare-earth compounds near an antiferromagnetic instability. As
a result, a major thrust in the quest for a theoretical framework of critical
phenomena beyond the Landau-Ginzburg-Wilson paradigm has centered around the
question of how Kondo screening can be destroyed at a $T=0$ magnetic-ordering
transition.

An extensively discussed alternative to the spin-density-wave picture is local
quantum criticality,\cite{Coleman.01,Si.01} in which Kondo screening itself
becomes critical precisely at the magnetic-ordering transition. In the vicinity
of such a Kondo-destruction quantum critical point (QCP), scaling properties of
the order parameter are very different from those predicted by spin-density-wave
theory. For example, inelastic neutron-scattering experiments have shown that
the dynamical spin (order-parameter) susceptibility exhibits
frequency-over-temperature (or $\omega/T$) scaling, and displays a fractional
exponent in the frequency and temperature dependences over a wide-range of
momentum space.\cite{Schroeder.00}
In addition, Hall measurements have provided evidence that the single-particle
relaxation rate is linear in temperature,\cite{Friedemann.09} a behavior that
arises naturally from $\omega/T$ scaling of the single-particle Green's
function.

A microscopic model of local quantum criticality in heavy-fermion metals has
been provided\cite{Si.01,Si.03} through extended dynamical
mean-field theory, which maps the Kondo lattice model to a single magnetic
impurity coupled to both a fermionic conduction band and to one or more bosonic
baths representing collective spin fluctuations. The critical point in the
lattice model is then captured in terms of a critical destruction of the Kondo
effect in the single-impurity problem. Such a construction gives rise to a QCP
that is spatially local but has interacting critical modes in imaginary
time, and that correctly captures $\omega/T$ scaling in the order-parameter
susceptibility. While the full extended dynamical mean-field treatment simulates
the lattice through self-consistent determination of the band and bath densities
of states, valuable insight can be gained by studying Kondo-destruction QCPs in
pure impurity problems.

This work investigates the quantum-critical destruction of the Kondo effect in
pseudogap variants of the Ising-symmetric (or easy-axis) Bose-Fermi Anderson
(BFA) and Bose-Fermi Kondo (BFK) models. In each model, a local degree of
freedom couples to a band of conduction electrons having a density of states
that vanishes as $|\epsilon|^r$ on approach to the Fermi energy ($\epsilon=0$).
The local degree of freedom is also coupled via the $z$ component of its spin
to a bosonic bath having a density of states proportional to $\omega^s$ for
frequencies up to some cutoff $\omega_c$ (i.e., for $0<\omega<\omega_c$).

In the presence of a metallic conduction band (corresponding to an exponent
$r=0$), the BFA and BFK models with a sub-Ohmic bosonic bath characterized
by a bath exponent $0<s<1$ feature a second-order quantum phase transition
(QPT)\cite{Smith.96,Smith.99,Sengupta.00,SiSmithIngersent.99,Zhu.02,Zarand.02,%
Glossop.05,Glossop.07} between a Kondo-screened (fermionic strong-coupling) phase
and a localized (Kondo-destroyed) phase in which the bosons asymptotically
suppress spin-flip scattering and a residual impurity moment survives to $T=0$.
For $1/2<s<1$, the Ising-symmetry BFA and BFK models are
thought\cite{Zhu.02,Zarand.02,Glossop.05,Glossop.07} to share the same critical
properties as the corresponding sub-Ohmic spin-boson model, which features an
interacting QCP characterized by critical exponents that vary continuously with
the bath exponent $s$.

In the absence of the bosonic bath, by contrast, the pseudogap BFA and pseudogap
BFK models reduce, respectively, to the pseudogap Anderson and pseudogap Kondo
models, in which the depression of the low-energy density of states impedes the
formation of a many-body Kondo screening cloud and gives rise to
QPTs\cite{Withoff.90,Bulla.97,Buxton.98,Ingersent.02,Fritz.04,Glossop.11}
between a Kondo phase and a localized phase in which the impurity
exhibits a free spin-$\half$ at $T=0$. The QPTs in the
pseudogap Anderson and Kondo models belong to the same universality
class\cite{Buxton.98,Ingersent.02,Fritz.04,Glossop.11} and are described
by critical exponents that vary with the band exponent $r$.  The pseudogap
versions of the BFA and BFK models offer fascinating possibilities for
nontrivial interplay between two different mechanisms for destruction of the
Kondo effect; an interplay that we investigate in this work.

An SU(2)-symmetric version of the pseudogap BFK model (in which the Cartesian
components of the impurity spin couple to different bosonic baths sharing the
same exponent $s$) has been studied via perturbative renormalization-group
(RG) methods,\cite{Vojta.03,Kircan.04} while the Ising-symmetric case of
bosonic coupling to the $z$ component of the impurity spin has been the subject
of preliminary investigation using the numerical renormalization group
(NRG).\cite{Glossop.08} A spinless variant of the model (coupling the
impurity charge to a bosonic bath) has been addressed by perturbative and
numerical RG techniques.\cite{Chung.07} In all these previous studies, the
order-parameter susceptibility has been found to exhibit an anomalous $T^{-x}$
variation in the quantum-critical regime with an exponent $x=s$ independent of
the band exponent $r$. One of the objectives of the present work is to
investigate more carefully the universality of this observation.

Another motivation for the present study is to expand our understanding of the
conditions that lead to $\omega/T$- or dynamical scaling of critical
correlation functions near unconventional quantum criticality.
It is well known that a quantum impurity model with a bulk component that is
conformally invariant can be described by a boundary conformal field
theory.\cite{Affleck.91,Cardy.84} In any such theory, a conformal mapping can
be used to obtain correlators at temperatures $T>0$ from their $T=0$
counterparts. In particular, the zero-temperature two-point correlator of a
primary conformal field $\Phi$ with scaling dimension $\lambda$ exhibits a
power-law decay $\langle\Phi(\tau,T=0)\:\Phi(0,T=0)\rangle\sim\tau^{-2\lambda}$.
This gives rise, via a conformal mapping,
to the scaling form\cite{Ginsparg:89,Tsvelik:96,units}
\begin{equation}
\label{scaling_form}
\langle \Phi(\tau,T)\:\Phi(0,T) \rangle
  \sim \left(\frac{\pi T}{\sin(\pi \tau T)}
\right)^{2\lambda}.
\end{equation}
The Fourier transform of Eq.\ \eqref{scaling_form} can be performed
analytically and yields a function of $\omega_n/T$ that can be analytically
continued to real frequencies provided\cite{note1} that $2\lambda<1$.
Therefore, demonstration that a system obeys Eq.\ \eqref{scaling_form}
with $2\lambda<1$ is sufficient to show the presence of $\omega/T$ scaling,
a characteristic feature of interacting QCPs.

In the pseudogap Anderson model, the conduction-band density of states breaks
conformal invariance, while in the metallic ($r=0$) BFK model, the conduction band is
conformally invariant but the bosonic bath is not.
Nonetheless, local correlators of each model have been
found\cite{Kirchner.08,Glossop.11,Pixley.11} to exhibit a boundary conformal
scaling form in imaginary time, consistent with Eq.\ \eqref{scaling_form};
such properties have been attributed to an enhanced symmetry at the QCPs.
It is an interesting question whether the scaling form Eq.\ \eqref{scaling_form}
also applies to the critical correlators of the pseudogap BFA and BFK models,
where the densities of states of the fermionic and bosonic baths both
break conformal invariance in the bulk.

We address these issues through a combination of techniques. For the
particle-hole-symmetric pseudogap BFA model, we use a continuous-time quantum
Monte Carlo (CT-QMC) method,\cite{Gull.11,Pixley.10} which stochastically
samples a perturbation expansion in the Anderson hybridization, to probe static
and dynamical quantities. We also use the Bose-Fermi extension of the
NRG method\cite{Glossop.05,Glossop.07} to resolve the critical spectrum and
extract static critical exponents of the pseudogap BFK model. Our results show
that the two models are in the same universality class.

We find that within different ranges of the exponents $r$ and $s$, the measured
critical exponents are determined by the fermionic band alone, by the bosonic
bath alone, or by both the fermions and the bosons. We show that both the
single-particle Green's function and the local spin susceptibility obey the
scaling form of Eq.\ \eqref{scaling_form} with exponents $2\lambda < 1$,
proving that each correlator obeys $\omega/T$ scaling in both the quantum
coherent ($\omega > T$) and the relaxational ($\omega < T$) regimes. Agreement
with NRG results for static quantities confirms the ability of the CT-QMC
approach to study quantum-critical properties of models involving bosons.

The remainder of the paper is organized as follows:
Section \ref{sec:models} introduces the pseudogap Bose-Fermi models and
Sec.\ \ref{sec:methods} briefly describes the numerical methods used to solve
these models. An overview of the phase diagram in Sec. \ref{sec:phases}
is followed in Sec.\ \ref{sec:critical} by a detailed description of the
quantum-critical properties. These results are summarized and discussed in Sec.\
\ref{sec:summary}.

\section{Models}
\label{sec:models}

The Bose-Fermi Anderson impurity model with Ising-symmetric bosonic coupling
is described by the Hamiltonian
\begin{align}
\label{H_BFA}
H_{\mathrm{BFA}}
&= \sum_{\kk\sigma} \epsilon_{\kk} c_{\kk\sigma}^{\dag} c_{\kk\sigma}^{\pdag}
 + \epsilon_d (n_{d\uparrow} + n_{d\downarrow}) + U n_{d\uparrow}n_{d\downarrow}
 \notag \\
&+ \frac{V}{\sqrt{N_k}} \sum_{\kk, \sigma} \bigl( d_{\sigma}^{\dag}
 c_{\kk\sigma}^{\pdag} + c_{\kk\sigma}^{\dag}d_{\sigma}^{\pdag} \bigr) \\
&+ \sum_{q} \omega_q \phi_q^{\dag}\phi_q^{\pdag} + \half g(n_{d\uparrow} -
 n_{d\downarrow})\sum_q \bigl(\phi_q^{\dag} + \phi_{-q}^{\pdag}\bigr), \notag
\end{align}
where $c_{\kk\sigma}$ annihilates a conduction-band electron with wave vector
$\kk$, energy $\epsilon_{\kk}$, and spin $z$ component $\half\sigma$ with
$\sigma=1$ (or $\uparrow$) or $-1$ (or $\downarrow$); $d_{\sigma}$ annihilates
an impurity electron with energy $\epsilon_d$ and spin $z$ component
$\half\sigma$; $n_{d\sigma}=d_{\sigma}^{\dag}d_{\sigma}^{\pdag}$; $\phi_q$
annihilates a boson of energy $\omega_q$; and $N_k$ is the number of unit
cells in the host (i.e., the number of distinct $\kk$ points). The other
energy scales entering Eq.\ \eqref{H_BFA} are the Coulomb repulsion $U$
between two electrons in the impurity level, the local hybridization $V$
between the impurity level and the conduction band, and the coupling $g$
between the $z$ component of the impurity spin and the bosonic bath.
We focus in this paper on cases $\epsilon_d=-\frac{1}{2}U$ corresponding to
particle-hole-symmetric impurities, but briefly discuss the effect of breaking
this symmetry in Sec.\ \ref{subsec:p-h-asymm}.

Over a wide region of its parameter space, the low-energy properties of
$H_{\mathrm{BFA}}$ can be mapped\cite{Pixley.10} via a Schrieffer-Wolff
transformation onto the Ising-symmetry Bose-Fermi Kondo Hamiltonian:
\begin{multline}
\label{H_BFK}
H_{\text{BFK}}
= \sum_{\kk\sigma} \epsilon_{\kk} c_{\kk\sigma}^{\dag} c_{\kk\sigma}^{\pdag}
   + J \, \mathbf{S}\cdot \mathbf{s}_c + \frac{W}{N} \sum_{\kk,\kk'\sigma}
   c_{\kk\sigma}^{\dag} c_{\kk'\sigma}^{\pdag} \\
+ \sum_q \omega_q \phi_q^{\dag}\phi_q^{\pdag}
   + g S_z \sum_q \bigl(\phi_q^{\dag} + \phi_{-q}^{\pdag}\bigr) ,
\end{multline}
where $J$ is the Kondo coupling, $W$ parameterizes nonmagnetic potential
scattering, and
\begin{equation}
\mathbf{s}_c = \frac{1}{2N_k} \sum_{\kk,\kk'\sigma,\sigma'}
   c_{\kk\sigma}^{\dag} \bm{\sigma}_{\sigma \sigma'} c_{\kk'\sigma'}^{\pdag}
\end{equation}
is the on-site conduction-band spin, with $\bm{\sigma}$ being a vector of Pauli
spin matrices. For $\epsilon_d=-\half U$, the potential scattering vanishes
for electrons on the Fermi surface, and we can set $W=0$. Although the bare Kondo
exchange term is SU(2) symmetric, the bosonic coupling breaks spin rotational
invariance. As a result, the RG description of the BFK model in terms of
renormalized couplings requires consideration of an anisotropic exchange
$J_z S_z s_{c,z} + \half J_{\perp} (S^+ s_c^- + S^- s_c^+)$.

For both models, we assume a conduction-band (fermionic-bath) density of states
\begin{equation}
\label{rho_F}
\rho_F(\epsilon) = N^{-1} \sum_{\kk} \delta(\epsilon - \epsilon_{\kk})=
\rho_0 |\epsilon/D|^r \, \Theta(D - |\epsilon|)
\end{equation}
with a power-law pseudogap described by $0<r<\half$, and a sub-Ohmic
bosonic bath specified by
\begin{equation}
\label{rho_B}
\rho_B(\omega) = \sum_q \delta(\omega - \omega_q)
  = K_0^2 \, \omega_c^{1-s} \omega^s \Theta(\omega) \Theta(\omega_c - \omega)
\end{equation}
with $\half\le s<1$.
The pseudogap density of states $\rho_F(\epsilon)$ leads to a BFA model with the
hybridization function $\Gamma_F(\epsilon)= \pi V^2 \rho_F(\epsilon) =
\Gamma\left|\epsilon/D\right|^r \Theta(D-|\epsilon|)$, where
$\Gamma = \pi \rho_0 V^2$.

Various limiting cases of Eqs.\ \eqref{H_BFA} and \eqref{H_BFK} have been
studied previously. For $g=0$, the pseudogap Bose-Fermi models simplify to
their pure-fermionic counterparts, in which a pseudogap critical point
separates Kondo and free-moment phases; at particle-hole symmetry, this
critical point exists only for $0<r<\half$ (Ref.\ \onlinecite{Buxton.98}).
In the absence of the conduction band, $H_{BFA}$ and $H_{BFK}$ both reduce
to the sub-Ohmic spin-boson model in zero transverse field, which has two
degenerate ground states in which the bosonic coupling localizes the impurity
either in its up- or down-spin configuration.
For $r=0$, the pseudogap BFA and BFK models reduce to their metallic
counterparts where the critical properties for Ising symmetry are
thought\cite{Zhu.02,Zarand.02,Glossop.05,Glossop.07} to coincide with those of
the spin-boson model for $\half\le s<1$.
In the sections that follow we show that the quantum criticality in the \emph{full}
pseudogap Bose-Fermi models described by Eqs.\ \eqref{H_BFA} and \eqref{H_BFK}
falls into one of three distinct types, depending on the values of the bath
exponents $r$ and $s$, with one of these types being governed by a mixed
Bose-Fermi QCP unlike those seen in any of the limiting cases.

\section{Methods}
\label{sec:methods}

\subsection{Continuous-time quantum Monte Carlo}

This work uses a form of the CT-QMC method described in Ref.\
\onlinecite{Pixley.10}. To make $H_{\text{BFA}}$ suitable for application
of the approach, we apply a canonical transformation to eliminate the term
linear in bosonic operators. This is achieved by the generator $S =
\frac{1}{2}g(n_{d\uparrow} - n_{d\downarrow})\sum_q \omega_q^{-1}(\phi_q^{\dag}
- \phi_{-q}^{\pdag})$, which transforms Eq.\ \eqref{H_BFA} to
\begin{align}
\tilde{H}_{\text{BFA}}
&= e^S H_{\text{BFA}} e^{-S} \notag \\
&= \sum_{\kk\sigma} \epsilon_{\kk} c_{\kk\sigma}^{\dag}c_{\kk\sigma}^{\pdag} +
 \tilde{\epsilon}_d(\tilde{n}_{\uparrow} + \tilde{n}_{\downarrow})
 + \tilde{U} \tilde{n}_{\uparrow}\tilde{n}_{\downarrow} \notag \\
&+ \frac{V}{\sqrt{N_k}} \sum_{\kk\sigma}\bigl(\tilde{d}_{\sigma}^{\dag}
 c_{\kk\sigma}^{\pdag} + c_{\kk\sigma}^{\dag}\tilde{d}_{\sigma}^{\pdag}\bigr)
 + \sum_q \omega_q \phi_q^{\dag}\phi_q^{\pdag} ,
\end{align}
where $\tilde{d}_{\sigma}^{\dag}= d_{\sigma}^{\dag} \exp[\frac{1}{2}\sigma
g\sum_q\omega_q^{-1}(\phi_q^{\dag} - \phi_{-q}^{\pdag})]$, $\tilde{U} = U +
\frac{1}{2}g^2\sum_q\omega_q^{-1}$, and $\tilde{\epsilon}_d=-\half\tilde{U}$.
Physically, the canonical transformation can be viewed as dressing the impurity
with a bosonic cloud, which in turn renormalizes $U$, $\epsilon_d$,
$d_{\sigma}$, and $d_{\sigma}^{\dag}$ without breaking particle-hole symmetry.

We are now in a position to calculate the partition function by expanding in
the hybridization $V$. This is similar to the approach in Refs.\
\onlinecite{Werner.07} and \onlinecite{Werner.10}, except that we couple the
bosonic bath to the $z$ component of the impurity's spin rather than to its
occupancy. [An extension of this method to the case of an SU(2)-symmetric
spin-boson coupling has recently been proposed.\cite{Otsuki.13}]
The resulting perturbation expansion is then sampled stochastically using a
Metropolis algorithm. Tracing out the fermionic band, which is unchanged by
the presence of bosons,\cite{Gull.11} allows the partition function to be
recast as $Z=\sum_k \int D[k] \, W_k$, where at perturbation order
$k$ there are $k$ segments along the imaginary time axis, each one defined by a
pair of renormalized impurity operators, $\tilde{d}_{\sigma'_j}(\tau'_j)$ and
$\tilde{d}^{\dag}_{\sigma'_j}(\tau''_j)$. The weight of a particular
configuration of segments is $W_k = w_F w_{\text{loc}} w_B$,
where $w_F$ is the weight of the band fermions, $w_{\text{loc}}$ is the
weight of the local configuration and $w_B$ is the weight of the bosonic bath.
$w_F$ can be obtained as a product of determinants and $w_{\text{loc}}$ is
calculated in terms of the length and overlap of imaginary time segments; for
details see Ref.\ \onlinecite{Gull.11}. If the operators in a given segment
configuration are time-ordered, with $s_i = 1$ ($-1$) indicating that the
$i^{\text{\,th}}$ operator acts at time $\tau_i$ to create (annihilate) an
electron of spin $z$ component $\sigma_i$, then the bosonic weight is
$w_B=\langle e^{\frac{1}{2}gs_{2k}\sigma_{2k} \hat{B}(\tau_{2k})} \cdots
e^{\frac{1}{2}gs_{1}\sigma_{1} \hat{B}(\tau_{1})} \rangle$, where $\hat{B}(\tau)
= \sum_q \omega_q^{-1} [\phi_q^{\dag}(\tau) - \phi_{-q}^{\pdag}(\tau)]$.
Tracing out the bosons and performing the momentum summation yields
\begin{equation}
w_B = \exp \Biggl\{-\frac{g^2}{4}\Biggl[ k B(0) + \sum_{1 \leq i<j \leq 2k}
s_i s_j \sigma_i \sigma_jB( \tau_j -\tau_i)\Biggr] \Biggr\}.
\label{w_B}
\end{equation}
To calculate the function $B(\tau)$ we replace the hard cut-off of the bosonic
spectral function in Eq.\ \eqref{rho_B} by an exponential cut-off,
$\rho_B(\omega) \to K_0^2 \omega_c^{1-s} \omega^s \Theta(\omega)
\exp(-\omega/\omega_c)$, which yields

\begin{align}
\label{B(tau)}
B(\tau)
&= K_0^2 \, (\omega_c/T)^{1-s} \, \Gamma(s-1)
   \Big[\zeta\bigl(s-1,\tau T + T/\omega_c\bigr) \notag \\
& \;\;\; +\zeta\bigl(s-1,1-\tau T + T/\omega_c\bigr)\Big],
\end{align}
where $\Gamma(x)$ is the gamma function and $\zeta(t,z)$ is the Hurwitz zeta
function.\cite{Weiss}  Now the local update procedure is identical to that in
Ref.\ \onlinecite{Werner.07}. In all the CT-QMC calculations reported below,
we have considered the low-temperature scaling limit ($T \ll \omega_c$) and
consequently dropped the terms $T/\omega_c$ from Eq.\ \eqref{B(tau)}. We have
checked that retaining or discarding these terms does not change any of the
universal features at the QCP such as critical exponents.

We perform an unbiased CT-QMC study of the quantum-critical properties of the
pseudogap BFA model by applying finite-temperature scaling to determine the
location of the QCP. We measure powers of the local magnetization
\begin{equation}
\langle M_z^{\,n} \rangle = \biggl\langle\biggl[
   T\int_0^{1/T}\,d\tau S_z(\tau) \biggr]^n \biggr\rangle,
\end{equation}
where $S_z(\tau) = \frac{1}{2}[n_{d\uparrow}(\tau) - n_{d\downarrow}(\tau)]$,
from which the Binder cumulant
\begin{equation}
\label{Binder}
U_4(T,g) = \displaystyle\frac{\langle M_z^4\rangle}{\langle M_z^2\rangle^2}
\end{equation}
is obtained. Similar to the case without bosons\cite{Glossop.11,
Pixley.11} we find that swap moves\cite{Gull.11} are essential to accurately
calculate the Binder cumulant.  Near classical phase transitions, the Binder
cumulant is a function of system size and temperature, and finite-size scaling
allows one to obtain the critical temperature in the thermodynamic limit from
classical Monte Carlo simulations of finite systems.\cite{Binder.81} For the
current quantum-mechanical problem, the inverse temperature $1/T$ and
coupling $g$ play the roles of system size and temperature, respectively, so
taking the zero-temperature limit allows one to determine the critical coupling
$g_c$. In the vicinity of the critical point, the Binder cumulant obeys the
scaling form\cite{Binder.81}
\begin{equation}
\label{Binder:scaling}
U_4(T,g) = \tilde{U}_4 \biggl(\frac{g/g_c-1}{T^{1/\nu}}\biggr) ,
\end{equation}
where $\nu$ is the correlation-length exponent defined in Eq.\ \eqref{nu}
below. Therefore, finite-temperature scaling applied to the Binder cumulant
can be used to extract the correlation-length critical exponent.

We also use CT-QMC to measure the single-particle Green's function
$G_{\sigma}(\tau, T) = \langle T_{\tau} d_{\sigma}(\tau) d_{\sigma}^{\dag}(0)
\rangle$ [$\equiv G(\tau,T)$ in zero magnetic field]
and the local spin susceptibility $\chi_{\text{loc}}(\tau, T) =
\langle T_{\tau} S_{z}(\tau) S_{z}(0) \rangle$, as described in Ref.\
\onlinecite{Gull.11}. For the susceptibility calculations presented in this
paper, we have used the segment representation, and have checked that the
results are consistent with those calculated using a ``matrix''
formalism.\cite{Gull.11} The static local spin susceptibility can then be
determined as $\chistat(T) = \chi_{\text{loc}}(\omega=0,T) = \int_0^{1/T}
d \tau \, \chi_{\text{loc}}(\tau, T)$, where we have set the Land\'{e} g factor
and the Bohr magneton to unity. For the noninteracting problem ($U=g=0$), with
$r=0.4$, we find agreement within numerical accuracy between the exact and
CT-QMC results for both the Green's function and the local spin
susceptibility.\cite{Pixley.10}

The CT-QMC results reported below were obtained for $r=0.4$,
$\omega_c=K_0^{-1}=D$, $\Gamma=0.1D$, $U=0.01D$, and for two different bath
exponents: $s=0.6$ and $s=0.8$. It is known from previous work\cite{Glossop.11}
that in the absence of bosons, the QCP point for $r=0.4$ and $\Gamma=0.1D$
occurs at $U_c \simeq 0.085D$. Therefore, by fixing $U=0.01D$ and adjusting the
bosonic coupling $g$, we are able to disentangle the pseudogap Bose-Fermi QCP
from the pure-fermionic pseudogap QCP.

\subsection{Numerical renormalization group}

The Bose-Fermi NRG\cite{Glossop.05,Glossop.07} treatment of the Hamiltonian
\eqref{H_BFK} entails three key steps: (i) Partition of the  fermionic and
bosonic baths described by $\rho_F(\epsilon)$ and $\rho_B(\omega)$ into
logarithmic bins spanning the energy ranges $\Lambda^{-j} < |\epsilon|/D,
\; \omega/\omega_c \le\Lambda^{-(j-1)}$, where $j = 1$, 2, 3, $\ldots$ and
$\Lambda > 1$ is the Wilson discretization parameter.
Within each logarithmic bin, the continuum is replaced by a single state,
namely, the linear combination of states that couples to the impurity.
(ii) Tridiagonalization of the baths, yielding the Hamiltonian
\begin{align}
\label{H_BFK:mapped}
H_{\text{BFK}}
&= D \sum_{n=0}^{\infty} \sum_{\sigma} \bigl[ \epsilon_n f_{n\sigma}^{\dag}
 f_{n\sigma}^{\pdag} + \tau_n \bigl( f_{n\sigma}^{\dag} f_{n-1,\sigma}^{\pdag}
 + \text{H.c} \big)\bigr] \notag \\
&+ \omega_c \sum_{m=0}^{\infty} \bigl[ e_m b_m^{\dag} b_m^{\pdag}
 + t_m \bigl( b_m^{\dag} b_{m-1}^{\pdag} + \text{H.c} \bigr)\bigr] \\
&+ F^2 \rho_0 J \vec{S} \cdot \sum_{\sigma,\sigma'} f^{\dag}_{0\sigma}
 \vec{\sigma}_{\sigma \sigma'} f_{0\sigma'}^{\pdag}
 + B K_0 g S_z \bigr( b_0^{\pdag} + b_0^{\dag} \bigr) . \notag
\end{align}
All information about $\rho_F(\epsilon)$ is encoded in $D$ and the
dimensionless coefficients $F$ and $\{\epsilon_n, \tau_n\}$, while
$\rho_B(\omega)$ is parametrized by $\omega_c$ and the dimensionless quantities
$B$ and $\{e_m, t_m\}$. For a particle-hole-symmetric $\rho_F(\epsilon)$ such
as that in Eq.\ \eqref{rho_F}, $\epsilon_n = 0$ for all $n$. For large values
of of $n$, the remaining tight-binding coefficients satisfy
$\tau_n \sim D\Lambda^{-n/2}$ and $|e_m|, \, t_m \sim \omega_c\,\Lambda^{-m}$.
(iii) Iterative solution of Eq.\ \eqref{H_BFK:mapped} on fermionic chains
restricted to sites $0\le n \le N$ with $N = 0$, 1, 2, $\ldots$. Due to the
faster decay of the bosonic hopping coefficients with increasing index $m$, the
bosonic chain is restricted to $0\le m\le N/2$, meaning that a site is added to
this chain only at even values of $N$. As in the conventional (pure-fermionic)
NRG, the $N_s$ many-body states of lowest energy are retained to form the basis
for iteration $N+1$. In problems involving bosonic baths, it is also necessary
to truncate the Fock space on each site of the bosonic chain. In this work we
employed a basis of boson number eigenstates $0 \le b_m^{\dag} b_m \le N_b$.

\begin{figure}[t]
\includegraphics[height=1.8in]{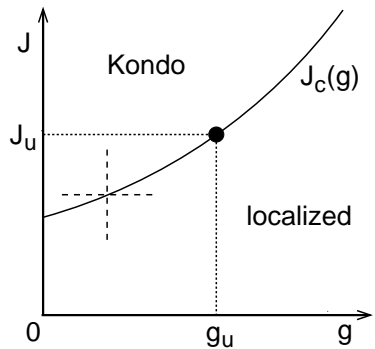}
\caption{\label{fig:phase_diagram} 
Schematic zero-temperature phase diagram for the pseudogap BFK model on the
plane spanned by the bosonic coupling $g$ and the Kondo exchange coupling $J$.
The system undergoes a QPT at any crossing of the solid line $J=J_c(g)$
representing the boundary between the two stable phases: Kondo and localized.
In this work, we consider crossings achieved by varying just one model
parameter, as exemplified by the horizontal and vertical dashed lines. The
labels $g_u$ and $J_u\equiv J_c(g_u)$ mark the point along the phase boundary
at which the temperature $T_u$, the upper limit of the quantum-critical regime
(see Fig.\ \ref{fig:crossover_scales}), takes its maximum value. The terminus
of the phase boundary at $g=0$, $J=J_c(0)$ corresponds to the QPT of the
pure-fermionic Kondo model. The pseudogap BFA model has a very similar phase
diagram on the $g$-$\Gamma$ plane at any fixed, positive value of
$U=-2\epsilon_d$.}
\end{figure}

The NRG calculation of critical exponents in the vicinity of a QCP is carried
out as described previously for the metallic ($r=0$) BFK
model\cite{Glossop.05,Glossop.07} and as illustrated in a preliminary
publication on the pseudogap BFK model.\cite{Glossop.08} The local magnetization
(equivalent to $\langle M_z\rangle$ above) and the static local susceptibility
are evaluated as
\begin{equation}
M_{\loc}= \langle S_z\rangle, \quad
\chistat=\partial M_{\loc}/\partial \hloc,
\end{equation}
where $\hloc$ is a magnetic field that couples only to the impurity
through an additional Hamiltonian term $\hloc S_z$. All results
reported below were obtained for $\omega_c=\sqrt{\pi}/K_0=D=\rho_0^{-1}=1$,
using Wilson discretization parameter $\Lambda=9$ and a bosonic truncation
parameter $N_b = 8$, and retaining after each iteration $N_s = 500$ many-body
multiplets corresponding to approximately 900 eigenstates.\cite{multiplets}
Experience from previous studies\cite{Glossop.05,Glossop.07,Glossop.08}
indicates that critical exponents calculated for these NRG parameter choices are
well-converged with respect to discretization errors (induced by the departure
of $\Lambda$ from its continuum value 1) and truncation errors (arising from
restricting the values of $N_s$ and $N_b$).

Over the range of bosonic exponents $0<s<1$, the Bose-Fermi NRG yields
critical exponents for the case $r=0$ that reproduce those obtained via
NRG\cite{Vojta.05} for the spin-boson model having the same exponent $s$. It
has been suggested in Refs.\ \onlinecite{Vojta.10} and \onlinecite{Vojta.12}
that, for $0<s<\half$, errors associated with the NRG calculation on a finite
bosonic Wilson chain lead to unphysical results with nontrivial exponents and
hyperscaling. On the other hand, for this same range of $s$,
Refs.\ \onlinecite{Kirchner.09}--\onlinecite{Kirchner.12} have demonstrated an
$\omega/T$ scaling for both the leading and sub-leading components of the
self-energy, which provides evidence for the interacting nature of the fixed
point. Since this issue has yet to be fully resolved, we restrict ourselves in
this paper to study of the range $\half \le s <1$.

\section{Phases}
\label{sec:phases}

\begin{figure}[t]
\includegraphics[height=1.5in]{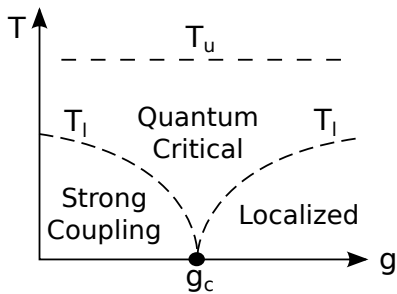}
\caption{\label{fig:crossover_scales} 
Schematic $T$ vs $g$ diagram in the vicinity of the pseudogap Bose-Fermi
QCP studied in this work. The QCP dominates the physics in a temperature
window between crossover scales $T_u$ and $T_l$ (dashed lines). For $g<g_c$ and
$0<T\lesssim T_l$, the system is in its Kondo regime, whereas for $g>g_c$
and $0<T\lesssim T_l$, the spin becomes decoupled from the conduction band
and the system is in the localized regime. At $T=0$ (only), cases $g<g_c$
and $g>g_c$ correspond to different phases, distinguished by the value of the
order parameter $\psi$ defined in Eq.\ \eqref{psi}.}
\end{figure}

In this section, we describe the two zero-temperature phases of the
pseudogap BFA and BFK models that can be accessed by tuning one model
parameter while holding all other parameters constant, as illustrated
in the case of the BFK model by the horizontal and vertical dashed lines in
Fig.\ \ref{fig:phase_diagram}. We will assume for the moment that the bosonic
coupling $g$ is the parameter that is varied.  
It
is important to emphasize
that, if carried out at any temperature $T>0$, such a variation invariably
produces smooth crossovers in physical properties, represented schematically
by the crossing of dashed lines on the $T$ vs $g$ diagram in Fig.\
\ref{fig:crossover_scales}. Only at $T=0$ is it possible to drive the
pseudogap BFA or BFK model through a QPT at $g=g_c$ separating a Kondo phase
(reached for $g<g_c$) from a localized phase (accessed for $g>g_c$).

The Kondo phase can be characterized by the vanishing (as $T\to 0$)
of the long-imaginary-time value of the local spin susceptibility
$\chi_{\text{loc}}(\tau=1/2T, T)$. Correspondingly, the static local
susceptibility $\chistat(T)$ approaches a constant at low temperatures (see,
for example, the squares in Figs.\ \ref{fig:stat6} and \ref{fig:stat8}),
signaling that the impurity spin is screened. Since these behaviors (and
all other universal low-energy properties found for $g<g_c$) prove to be
identical to those in the Kondo phase of the pseudogap Anderson and Kondo
models, and independent of the bosonic bath exponent $s$, we associate them
with an RG fixed point located at renormalized couplings $g=0$ and either
$\Gamma=\infty$ or $J=\infty$.

In the localized phase, by contrast, $\chi_{\text{loc}}(\tau=1/2T, T)$
approaches a constant $C(g)>0$ in the limit $T\to 0$ and the static local
susceptibility assumes the Curie-law form $\chistat(T)=C(g)/T$ (triangles in
Figs.\ \ref{fig:stat6} and \ref{fig:stat8}) characteristic of a  
free spin whose size is $\propto \sqrt{C(g)}$.
In this low-temperature regime, the impurity spin is
essentially decoupled from the conduction band and follows the fluctuations of
the bosonic bath. The asymptotic low-energy properties are governed by an RG
fixed point located at renormalized couplings $g=\infty$ and either $\Gamma=0$
or $J=0$.

The two phases described above are separated by a QPT occurring at bosonic
coupling $g_c$ where the Curie constant $C(g)$ extrapolates continuously to
zero as $g$ is decreased. As a result, $\lim_{T \rightarrow 0} T\chistat$ can
serve as an order parameter for the QPT,\cite{Buxton.98,Ingersent.02} vanishing
throughout the Kondo phase and equaling the Curie constant in the localized
phase. However, it is conventional instead to take as the order parameter
\begin{equation}
\label{psi}
\psi = \lim_{h_{\loc}\to \, 0^-} M_{\loc}(T=0),
\end{equation}
which rises continuously from zero on entry into the localized phase.

If one fixes all parameters apart from $\Gamma$ and $g$ in the BFA model
[$J$ and $g$ in the BFK model], then the function $g_c(\Gamma)$ [$g_c(J)$]
defines the boundary between the Kondo and localized phases. This boundary,
shown for the Kondo case as the solid line in Fig.\ \ref{fig:phase_diagram}, is
anchored at $g=0$ by the QCP of the pure-fermionic pseudogap models. One of the
central questions addressed in our work is whether the critical behavior reached
by crossing the phase boundary at $g>0$ coincides with or differs from that for
$g=0$.

Our two numerical techniques lend themselves to different approaches for
locating the phase boundary. In the CT-QMC treatment of the pseudogap BFA model,
the low-temperature limit of the Binder cumulant $U_4$ evolves continuously
from $U_4=3$ in the Kondo phase to $U_4=1$ in the localized phase. $U_4$ is
independent of temperature at the critical bosonic coupling $g_c$, but not at
other nearby values of $g$. Thus, one can find $g_c$ via the intersection of
curves $U_4$ vs $g$ for different (low) temperatures (see Figs.\
\ref{fig:binder6} and \ref{fig:binder8}).

Within the NRG, one can identify the phase boundary of the pseudogap BFK model
through examination of the asymptotic low-energy many-body spectrum, the
$T\to 0$ values of thermodynamic properties such as the impurity contribution
to the entropy ($S_{\text{imp}}= 2r\ln 2$ in the Kondo phase, and
$S_{\text{imp}}=\ln 2$ in the localized phase), or the static local spin
susceptibility $\chistat(T)$. In the present study of the pseudogap BFK model,
rather than calculating $g_c$ as a function of $J$, we have instead determined
the critical Kondo coupling $J_c$ for different values of $g$. For given
bath exponents $r$ and $s$, points on the phase boundary at any $g>0$ are all
found to share the same many-body spectrum and the same power laws in the
quantum-critical regime $0\le T\lesssim T_u$ (see Fig.\
\ref{fig:crossover_scales}). This universality strongly suggests that the
quantum criticality is governed by a single QCP to which the system flows
starting from any point on the curve $J_c(g)$.
However, the upper temperature $T_u$, which marks the energy scale at which
RG flow first brings the system under the influence of the QCP, varies widely
along the boundary. In order to provide the most accurate possible account of
the quantum-critical properties, we focus below on NRG results obtained for the
boundary point $g=g_u$, $J=J_u\equiv J_c(g_u)$ that yields the highest value of
$T_u$ for given $(r,s)$, and can therefore be assumed to lie closest to the QCP.
An advantage of this approach is that for certain $(r,s)$ pairs we find $g_u=0$,
making clear that in such cases the QCP is of pure-fermionic character.

In what follows, we write $\Delta=g-g_c$ for the BFA model and $\Delta=J_c-J$
for the BFK model to denote the system's distance from the phase boundary, with
$\Delta<0$ describing the Kondo phase and $\Delta>0$ describing the
localized phase.

\section{Results Near the Quantum Critical Point}
\label{sec:critical}

\subsection{Static Critical Behavior}

At any point lying on the phase boundary between the Kondo and
localized phases, the static susceptibility exhibits a temperature dependence
\begin{equation}
\label{chi_stat:crit}
\chistat(T;\Delta=0) \simeq A T^{-x(r,s)}
\end{equation}
for all temperatures in the quantum-critical window $0<T\lesssim T_u$. Here,
the exponent $x$ is a universal property of the pseudogap Bose-Fermi QCP (i.e.,
$x$ depends only on the bath exponents $r$ and $s$), whereas the crossover
temperature $T_u$ and the prefactor $A$ can vary from point to point along the
phase boundary; as noted above, $T_u$ is highest for the boundary point that
lies nearest to the QCP.

Close to but not precisely on the phase boundary, the system enters the
quantum-critical regime once the temperature drops below roughly the same scale
$T_u$ found for $\Delta=0$. However, further lowering the temperature produces a
second crossover to an asymptotic regime $T\lesssim T_l$ governed either by the
Kondo fixed point (for $\Delta<0$) or by the localized fixed point
(for $\Delta>0$); see Fig.\ \ref{fig:crossover_scales}. Unlike $T_u$, the lower
crossover scale $T_l$ shows significant $\Delta$ dependence, and vanishes
continuously upon approach to the phase boundary according to
\begin{equation}
\label{nu}
T_l \propto |\Delta|^{\nu} ,
\end{equation}
where $\nu$ is the correlation-length exponent.

In the absence of bosons, the exponent $x(r,s)$ necessarily assumes the value
$x_F(r)$ found at the QCP of the pure-fermionic pseudogap Anderson and Kondo
models. With a constant fermionic density of states (i.e., $r=0$) and
isotropic, XY, or Ising symmetry of the bosonic couplings, it is
known\cite{Zhu.02,Zarand.02} that $x(0,s)$ for $\half<s<1$ reduces to the
exponent $x_B(s)=s$ of the spin-boson model. Based on perturbative
RG,\cite{Vojta.03,Kircan.04} it has been concluded that $x(r,s)$ for a
spin-isotropic version of the pseudogap BFK model is independent of $r$, an
observation that agrees with asymptotically exact results obtained in the
dynamical large-$N$ limit\cite{Parcollet.98,Zhu.04} where the symmetry group
of the spin-isotropic pseudogap BFK model is generalized from SU(2) to
SU($N$).\cite{Zamani.12}
A preliminary NRG study\cite{Glossop.08} of the easy-axis pseudogap BFK model
also found $x(r,s)=s$. However, that study considered only exponent pairs
$(r,s)$ for which $x_B(s)<x_F(r)$, a regime in which it is quite plausible that
the bosons should dominate the singular part of the spin response. In the
present study we have also investigated cases where $x_F(r)<x_B(s)$
that offer better prospects for finding fermion-dominated spin dynamics.

In what follows we show using both CT-QMC calculations and the NRG that the
static magnetic critical exponent $x$ is simply the smaller of the exponents
governing the cases of pure-fermionic and pure-bosonic critical Kondo
destruction:
\begin{equation}
\label{x_BF}
x(r,s) = \mathrm{min}\bigl[x_F(r),x_B(s)\bigr].
\end{equation}
We focus primarily on the pseudogap exponent $r=0.4$ and two values of
the bath exponent: $s=0.6$ and $0.8$. For the pure-fermionic Kondo model,
the critical susceptibility is described by a temperature exponent
$x_F(r=0.4)=0.688(1)$ obtained within the NRG,\cite{Ingersent.02} consistent
with the value $x_F=0.68(3)$ found using CT-QMC for the corresponding Anderson
model.\cite{Glossop.11}
(Here and throughout the remainder of the paper, a number in parentheses
indicates our estimated uncertainty in the last digit.)
For the Bose-Fermi models, therefore, the cases
$(r,s)=(0.4,0.6)$ and $(r,s)=(0.4,0.8)$ are representative of the regimes
$x_B<x_F$ and $x_B>x_F$, respectively. For each of these cases, we present
CT-QMC results for the pseudogap BFA model and NRG results for the pseudogap
BFK model, finding the critical behavior of the two models to be fully
equivalent.

We supplement the detailed results for $r=0.4$, $s=0.6$ and $0.8$ with static
critical exponents for a larger set of $(r,s)$ pairs as obtained for the BFK
model using the NRG. In addition to $x$ and $\nu$ introduced in Eqs.\
\eqref{chi_stat:crit} and \eqref{nu}, we consider exponents $\beta$ and
$\delta$ defined via the relations\cite{Glossop.07}
\begin{align}
\label{beta}
M_{\loc}(\Delta>0;T=0,\hloc\to 0)& \propto \Delta^{\beta}, \\
\label{delta}
M_{\loc}(\hloc;\Delta=0,T=0)& \propto |\hloc|^{1/\delta}.
\end{align}
If the QPT occurs below its upper critical dimension,
yielding an interacting QCP, one expects the singular component of the free
energy to take the form
\begin{equation}
\label{scaling-ansatz}
F_{\text{crit}}
= T f\biggl(\frac{|\Delta|}{T^{1/\nu}}, \, \frac{|\hloc|}{T^{(1+x)/2}} \biggr),
\end{equation}
where $f$ is a scaling function. With this ansatz, the exponents $\beta$ and
$\delta$ are related to $\nu$ and $x$ by hyperscaling relations
\begin{align}
\label{beta:hyper}
\beta&  = \nu (1-x) / 2, \\
\label{delta:hyper}
\delta& = (1 + x) / (1 - x).
\end{align}
 
In each of the cases we have studied, the value of $x$ is consistent with
Eq.\ \eqref{x_BF}, and wherever we have tested them, the hyperscaling
Eqs.\ \eqref{beta:hyper} and \eqref{delta:hyper} are well obeyed. Based on
the value of $\nu(r,s)$ [or alternatively, $\beta(r,s)$]---as well as an
analysis of the many-body spectrum at the QCP---we are led to subdivide the
region of the $(r,s)$ plane in which $x(r,s)=x_B(s)$ into two parts: one
in which the fermions appear to play no role in the critical behavior,
and another in which bosonic and fermionic fluctuations combine to produce
critical behavior unlike that found in either the metallic ($r=0$) BFK model
or the pseudogap Kondo model.

\subsubsection{Results for $r=0.4$, $s=0.6$}

\begin{figure}[t]
\includegraphics[width=2.2in,angle=270]{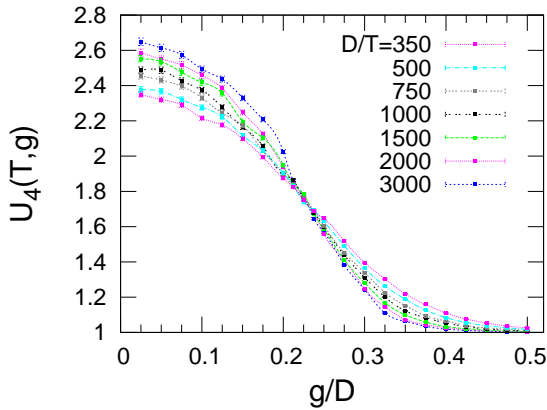}
\caption{\label{fig:binder6} (Color online)
Binder cumulant $U_4$ vs bosonic coupling $g$ for $r=0.4$ and $s=0.6$ at
different inverse temperatures $\beta=1/T$.
There is a clear intersection all curves at the critical coupling
$g_c/D = 0.225(7)$.
Error bars for Binder cumulant calculations are obtained from a jackknife
error analysis.}
\end{figure}

Figure \ref{fig:binder6} shows the variation with bosonic coupling $g$ of the
Binder cumulant $U_4$ of the pseudogap BFA model for $r=0.4$, $s=0.6$ at
different temperatures as calculated using CT-QMC. The intersection of the
curves places the QCP at $g_c/D=0.225(7)$.

\begin{figure}[t]
\includegraphics[width=2.5in,angle=270]{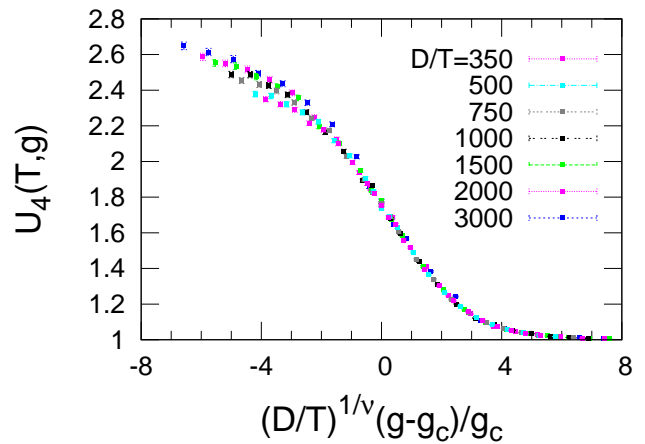}
\caption{\label{fig:nu6} (Color online)
Data collapse of the Binder cumulant $U_4$ for $r=0.4$ and $s=0.6$ using
Eq.\ \protect\eqref{Binder:scaling}. An inverse correlation-length exponent
$\nu(r=0.4,s=0.6)^{-1}=0.25(3)$ produces an excellent collapse of data
in the vicinity of the critical point.}
\end{figure}

The scaling form Eq.\ \eqref{Binder:scaling} of the Binder cumulant in the
vicinity of the QCP can be used to extract the correlation-length exponent
defined in Eq.\ \eqref{nu}. As illustrated in Fig.\ \ref{fig:nu6}, we obtain
an excellent collapse of data taken at different temperatures with a fitted
exponent $\nu(r=0.4,s=0.6)^{-1} = 0.25(3)$.

\begin{figure}[t]
\includegraphics[height=2.2in]{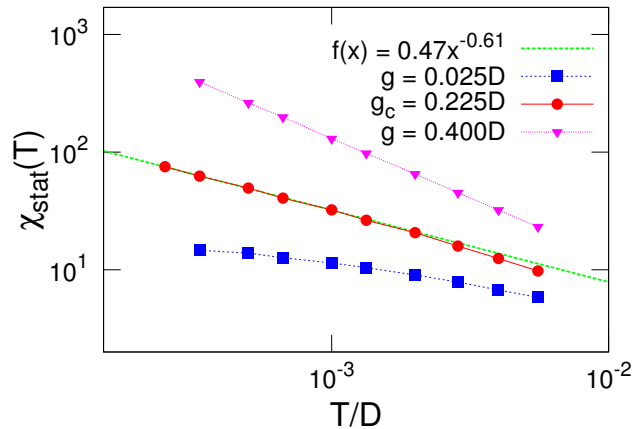}
\caption{\label{fig:stat6} (Color online)
Static spin susceptibility $\chistat$ from CT-QMC vs temperature $T$ for
$r=0.4$ and $s=0.6$ in the Kondo-screened phase (squares), at the critical
coupling (circles), and in the localized phase (triangles). At the critical
coupling $g_c\approx 0.225$, $\chistat$ diverges according to Eq.\
\eqref{chi_stat:crit} with $x=0.61(2)$.}
\end{figure}

Having found $g_c$, we are able to establish that the static local
susceptibility has the expected temperature dependence in each phase and at
the critical coupling, as illustrated in Fig.\ \ref{fig:stat6}. In
particular, for $g=g_c$, $\chistat$ follows
Eq.\ \eqref{chi_stat:crit} with $x(r,s)=0.61(2)$ over the lowest decade of
temperature for which data were obtained: $2.5\times 10^{-4}D\le T\lesssim
T_u\simeq 2.5\times 10^{-3}D$. (For comparison purposes, we note that for $g=0$,
the Kondo temperature is $T_K^0 = 0.06D$.)
To within numerical accuracy, we find that $x(r,s)=x_B(s)=s$, in agreement
with previous perturbative and numerical RG
studies.\cite{Vojta.03,Kircan.04,Glossop.08}

In the BFK model, the convergence of the NRG many-body spectrum to the
critical spectrum is fastest (i.e., the crossover scale $T_u$ is highest)
for $g_u/D=0.84(4)$, with $J_c(g=0.84D)/2D\simeq 0.9666$. (The value of
$J_c$ was determined to roughly 10 significant figures.)

\begin{figure}[t]
\includegraphics[height=2.2in]{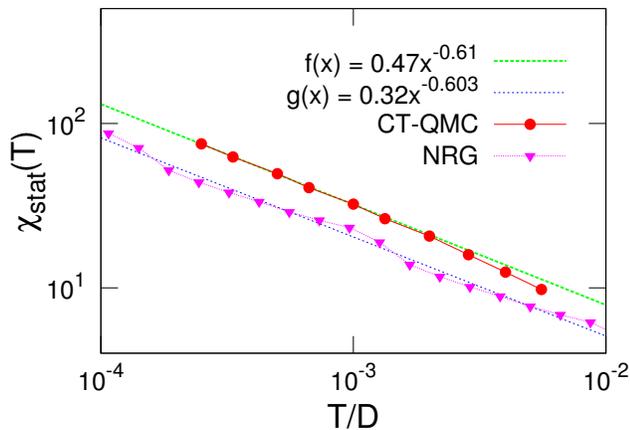}
\caption{\label{fig:stat6nrg} (Color online)
Static spin susceptibility $\chistat$ vs temperature $T$ for $r=0.4$ and
$s=0.6$ calculated within CT-QMC for the BFA model and using the NRG for the
BFK model at the QCP. The curves run parallel over the temperature range shown
and are fitted by consistent exponents.}
\end{figure}

Figure \ref{fig:stat6nrg} superimposes the temperature dependence of the
critical static local susceptibilities of the Anderson and Kondo models.
The NRG results for the Kondo model exhibit small oscillations around the
dependence predicted in Eq.\ \eqref{chi_stat:crit}. Such oscillations, which
are periodic in $\log T$ with period $\log \Lambda$, are a known
consequence of the NRG band discretization\cite{Oliveira.94} that can be reduced
in amplitude by working with smaller values of $\Lambda$. Over the two decades
of temperature shown in the figure, the NRG data are described by an exponent
$x\simeq 0.603$. However, a fit over the range $10^{-15} \le T/D \le 10^{-5}$
yields an improved estimate $x=0.600(1)$, consistent to within numerical
error with the CT-QMC result for the Anderson model. Like the CT-QMC estimate
for the Anderson model, this value is consistent with the hypothesis that for
$x_B(s)<x_F(r)$ spin fluctuations are primarily driven critical by the bosonic
bath, and $x(r,s)=x_B(s)=s$.

The reciprocal of the correlation-length exponent extracted from the crossover
in the NRG many-body spectrum is $\nu^{-1}=0.233(1)$, again consistent with the
value obtained for the Anderson model. This value differs from that found in
two other cases: in the metallic BFK model,\cite{Glossop.07}
$\nu(r=0,s=0.6)^{-1} = 0.509(1)$, while in the pseudogap Kondo
model,\cite{Ingersent.02} $\nu_F(r=0.4)^{-1} = 0.171(1)$. Although the critical
spin fluctuations (and hence the exponent $x$) for $(r,s)=(0.4,0.6)$ seem to be
dominated by the bosonic bath, the RG flow away from the critical point
described by the exponent $\nu$ is clearly different from that in cases of
pure-bosonic or pure-fermionic criticality.

\subsubsection{Results for $r=0.4$, $s=0.8$}

\begin{figure}[t]
\includegraphics[width=2.2in,angle=270]{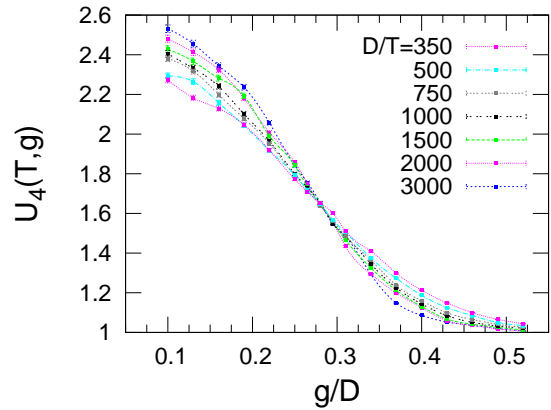}
\caption{\label{fig:binder8} (Color online)
Binder cumulant $B_4$ vs bosonic coupling $g$ for $r=0.4$ and $s=0.8$ at
different inverse temperatures $\beta=1/T$.
There is a clear intersection of the curves at the critical coupling
$g_c/D=0.28(1)$.}
\end{figure}

Figure \ref{fig:binder8} shows the variation of the BFA-model Binder cumulant 
for $r=0.4$, $s=0.8$ as calculated using CT-QMC. The intersection of curves
representing different temperatures places the QCP at $g_c/D=0.28(1)$.
This critical value is larger than that ($g_c/D\simeq 0.225$) for $s=0.6$ because
the bosonic interaction between time segments falls off faster with increasing
$s$. Figure \ref{fig:nu6} shows that the Binder cumulant scales according to
Eq.\ \eqref{Binder:scaling} with an excellent collapse of data taken at
different temperatures with a fitted exponent
$\nu(r=0.4,s=0.8)^{-1} = 0.17(2)$.

\begin{figure}[t]
\includegraphics[width=2.5in,angle=270]{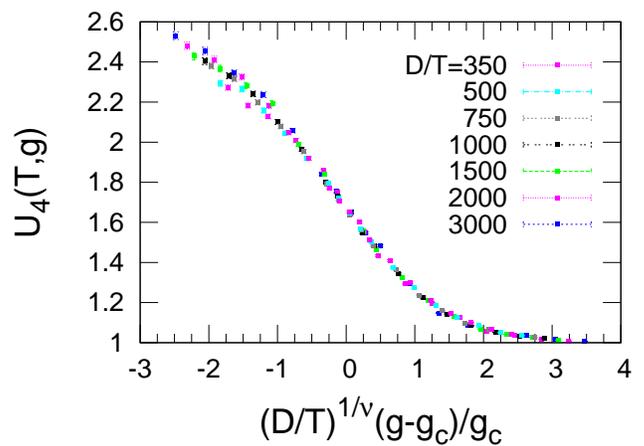}
\caption{\label{fig:nu8} (Color online)
Data collapse of the Binder cumulant for $r=0.4$ and $s=0.8$. A correlation
length exponent $\nu(r=0.4,s=0.8)^{-1}=0.17(2)$ produces an
excellent collapse in the vicinity of the critical point.}
\end{figure}

\begin{figure}[t]
\includegraphics[height=2.2in]{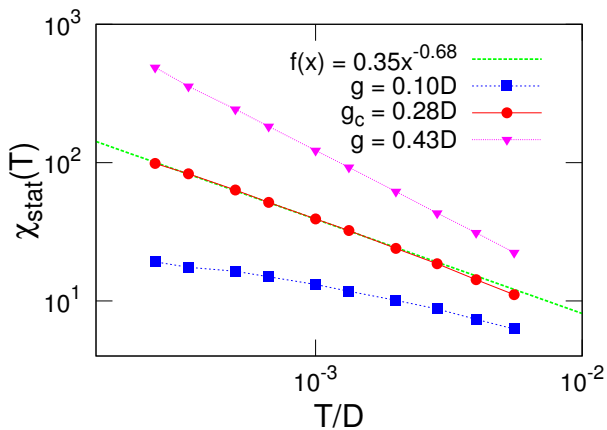}
\caption{\label{fig:stat8} (Color online)
Static spin susceptibility $\chistat$ from CT-QMC vs temperature $T$ for
$r=0.4$ and $s=0.8$ in the Kondo-screened phase (squares), at the critical
coupling (circles) and in the localized phase (triangles). At the critical
coupling $g_c/D=0.28(1)$, $\chistat$ diverges according to
Eq.\ \protect\eqref{chi_stat:crit} with $x=0.68(2)$.}
\end{figure}

Figure \ref{fig:stat8} illustrates the temperature dependence of the static
local susceptibility in each phase and at the critical coupling.
For $g=g_c$, $\chistat$ follows
Eq.\ \eqref{chi_stat:crit} with $x(r,s)=0.68(2)$ over the lowest decade of
temperature for which data were obtained, a clear departure from the
behavior $x(r,s)=x_B(s)=s$ seen above for $(r,s)=(0.4,0.6)$.

In the BFK model, the convergence of the NRG many-body spectrum to the
critical spectrum is fastest (i.e., the crossover scale $T_u$ is highest)
for $g_u=0$, with $J_u\equiv J_c(g_u)/2D\simeq 0.7908$. The value $g_u=0$
means that the impurity is entirely decoupled from the bosons, and the QCP
must be of pure-fermionic character. Indeed, the asymptotic low-energy
many-body spectrum in the quantum-critical regime can be reproduced by taking
every possible combination of (i) one state from the spectrum of free $s=0.8$
bosons, and (ii) one state from the critical spectrum of the $r=0.4$ pseudogap
Kondo model. We summarize this spectral decomposition in the shorthand
(BF critical) = (B free) $\otimes$ (F critical) and refer to it below simply
as F-type criticality.

\begin{figure}[t]
\includegraphics[height=2.2in]{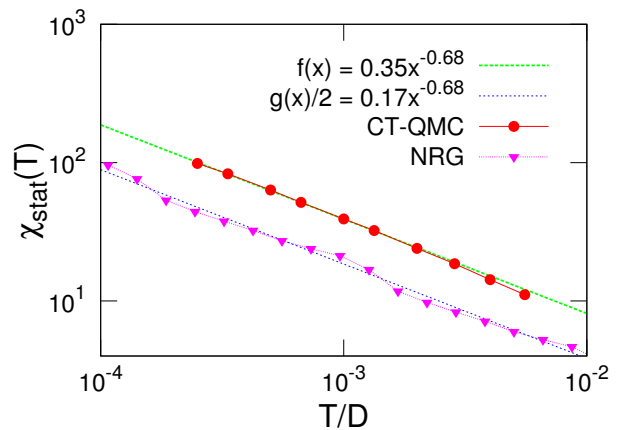}
\caption{\label{fig:stat8nrg} (Color online)
Static spin susceptibility $\chistat$ vs temperature $T$ for $r=0.4$ and
$s=0.8$ calculated within CT-QMC for the BFA model and using the NRG for the
BFK model at the QCP. The curves run parallel over the temperature range shown
and are described by consistent exponents.}
\end{figure}

Figure \ref{fig:stat8nrg} compares the temperature dependence of the
critical static local susceptibilities of the BFA and BFK models.
Over the two decades of temperature shown in the figure, the NRG data for the
BFK model are described by an exponent $x\simeq 0.68$, identical to the
CT-QMC value for the Anderson model. However, since the QCP corresponds to
$g=0$, we know that the exponent must coincide exactly with that
of the pseudogap Kondo model:\cite{Ingersent.02} $x_F(r=0.4)=0.688(1)$.
This value clearly differs from the one $x=s=0.8$ found within the
spin-boson-model\cite{Vojta.05} and the metallic BFK model.\cite{Glossop.05}

Similarly, we can be confident that the correlation-length exponent must be
identical to that of the pseudogap Kondo model, $\nu_F(r=0.4)^{-1}=0.171(1)$,
a result that is consistent with the CT-QMC value
$\nu(r=0.4,s=0.8)^{-1} = 0.17(2)$ quoted above.

The preceding results suggest that in cases $(r,s)$ where $x_B(s)>x_F(r)$,
the physics at and near the QCP is determined primarily by fermionic
fluctuations, and that critical properties should coincide with those of the
pseudogap Anderson and pseudogap Kondo models.

\subsubsection{Results for other $(r,s)$}
\label{subsubsec:other-r,s}

\begin{table*}
\begin{ruledtabular}
\begin{tabular}{dddcdddd}
&&& critical \\[-.5ex]
\multicolumn{1}{c}{$r$} &
\multicolumn{1}{c}{$s$} &
\multicolumn{1}{c}{$\;\;g_u/D$} &
spectrum &
\multicolumn{1}{c}{$\;\;x$} &
\multicolumn{1}{c}{$\;\;\beta$} &
\multicolumn{1}{c}{$\;\;1/\nu$} &
\multicolumn{1}{c}{$\;\;\;\;1/\nu_B(s)$}
\\[.5ex] \hline \\[-2ex]
0.1 & 0.5  & \multicolumn{1}{c}{$\quad> 10$}
		     & B & 0.499 &       & 0.48     & 0.475 \\
    & 0.7  & \multicolumn{1}{c}{\;\;6--8}
		     & B & 0.700 & 0.296 & 0.506\:* & 0.506 \\
    & 0.75 & \multicolumn{1}{c}{\;\;6--8}
		     & B &       & 0.254 & 0.493\:* & 0.493 \\
    & 0.8  & 5.0(5)  & B & 0.800 & 0.213 & 0.46     & 0.470 \\
    & 0.85 & 3.58(3) & M &       & 0.176 & 0.426\:* & 0.433 \\
    & 0.9  & 2.8(3)  & M & 0.900 & 0.142 & 0.352    & 0.376 \\
    & \multicolumn{1}{c}{no bosons} & \multicolumn{1}{c}{--} & --
			 & 0.989 & 0.060 & 0.093\:*
			 & \multicolumn{1}{c}{\;\;\;--} \\
\hline \\[-2.3ex]
0.2 & 0.5  & \multicolumn{1}{c}{$\quad> 10$}
		     & B & 0.499 & 0.526 & 0.48     & 0.475 \\
    & 0.6  & \multicolumn{1}{c}{\;\;\;\:5--10}
		     & B &       & 0.394 & 0.508\:* & 0.509 \\
    & 0.65 & 3.40(5) & M &       & 0.347 & 0.504\:* & 0.512 \\
    & 0.7  & 2.98(3) & M &       & 0.313 & 0.479\:* & 0.506 \\
    & 0.75 & 2.63(3) & M &       & 0.282 & 0.443\:* & 0.493 \\
    & 0.8  & 2.30(5) & M & 0.800 & 0.251 & 0.398    & 0.470 \\
    & 0.9  & 1.4     & M & 0.900 & 0.189 & 0.265\:* & 0.376 \\
    & \multicolumn{1}{c}{no bosons} & \multicolumn{1}{c}{--} & --
			 & 0.948 & 0.160 & 0.161\:*
			 & \multicolumn{1}{c}{\;\;\;--} \\
\hline \\[-2.3ex]
0.3 & 0.5  & 2.5     & M & 0.500 & 0.582 & 0.427(4) & 0.475 \\
    & 0.6  & 2.05(5) & M & 0.600 & 0.493 & 0.405(2) & 0.509 \\
    & 0.7  & 1.65(5) & M & 0.700 & 0.422 & 0.356\:* & 0.506 \\
    & 0.8  & 1.1     & M & 0.800 & 0.365 & 0.270    & 0.470 \\
    & 0.9  & 0       & F & 0.862\:\S & 0.355\:\S & 0.194\:{*\S} & 0.376 \\ 
    & \multicolumn{1}{c}{no bosons} & \multicolumn{1}{c}{--} & --
			 & 0.862 & 0.355 & 0.194\:*
			 & \multicolumn{1}{c}{\;\;\;--} \\
\hline \\[-2.3ex]
0.4 & 0.5  & 1.15(5) & M & 0.500 & 0.930       & 0.269    & 0.475 \\
    & 0.6  & 0.84(2) & M & 0.600 & 0.853       & 0.233    & 0.509 \\
    & 0.7  & 0       & F & 0.688\:\S & 0.914\:\S & 0.171\:{*\S} & 0.506 \\
    & 0.8  & 0       & F & 0.688\:\S & 0.914\:\S & 0.171\:{*\S} & 0.470 \\
    & 0.9  & 0       & F & 0.688\:\S & 0.914\:\S & 0.171\:{*\S} & 0.376 \\
    & \multicolumn{1}{c}{no bosons} & \multicolumn{1}{c}{--} & --
			 & 0.688 & 0.914 & 0.171\:*
			 & \multicolumn{1}{c}{\;\;\;--} \\
\end{tabular}
\end{ruledtabular}
\caption{\label{tab:NRG_exponents}
Summary of critical properties of the Bose-Fermi Kondo (BFK) model with
pseudogap exponent $r$ and bosonic bath exponent $s$. As defined in the text,
$g_u$ is the value of the bosonic coupling $g$ that yields the highest
temperature of entry into the quantum-critical regime. In the fourth column,
``F'' means that the asymptotic low-energy form of the critical spectrum is a
direct product of (i) the spectrum of free bosons with the same $s$ and (ii)
the critical spectrum of the pseudogap Kondo model with the same $r$; ``B''
indicates that the asymptotic low-energy NRG critical spectrum is a direct
product of (i) the critical spectrum of the spin-boson model with the same
value of $s$ and (ii) the Kondo spectrum of the pseudogap Kondo model with the
same $r$; and ``M'' means that the critical spectrum does not decompose into a
direct product of bosonic and fermionic parts. Exponents $x$, $\beta$, and
$\nu$ are as defined in Eqs.\ \eqref{chi_stat:crit}, \eqref{beta}, and
\eqref{nu}, respectively, while $1/\nu_B(s)$ is the reciprocal of the
order-parameter exponent at a pure-bosonic QCP with the same $s$, as calculated
in the metallic ($r=0$) BFK model. All values of $1/\nu_B$ and those values of
$1/\nu$ followed by an asterisk were obtained from the corresponding value of
$\beta$ using hyperscaling [Eq.\ \eqref{beta:hyper}] under the assumption
that $x=s$. Any exponent followed by ``$\S$'' has been set to that for the
pure-fermionic pseudogap Kondo model (see the line labeled ``no bosons'' for
each value of $r$) since $g_u=0$ indicates that the bosonic bath plays no part
in the criticality. A number in parentheses represents the uncertainty in the
last digit, equal to $1$ where omitted.}
\end{table*}

In order to investigate more systematically the different types of quantum
criticality exemplified in the cases $(r,s)=(0.4,0.6)$ and $(0.4,0.8)$, we have
studied the particle-hole-symmetric pseudogap BFK model for 23 different $(r,s)$
pairs spanning the ranges $0.1\le r\le 0.4$ and $0.5\le s\le 0.9$.
Table \ref{tab:NRG_exponents} summarizes the critical properties, one line
per $(r,s)$ pair. On each line, an estimate of $g_u$---the value of $g$ that
produces the highest temperature $T_u$ of entry into the quantum-critical
regime for $J\simeq J_u\equiv J_c(g_u)$, is followed by the assignment of the
critical many-body spectrum to one of three categories (F, B, or M) described
further below.

The remaining columns of Table \ref{tab:NRG_exponents} list critical exponents:
$x$, $\beta$, and $1/\nu$ [defined in Eqs.\ \eqref{chi_stat:crit}, \eqref{beta},
and \eqref{nu}, respectively] are values calculated or inferred for the pair
$(r,s)$ in question. For purposes of comparison, we also list $1/\nu_B(s)$, the
reciprocal of the order-parameter exponent at a pure-bosonic QCP with the same
$s$, as determined within the metallic BFK model.
The exponent $x$ has been calculated for all but a small number of $(r,s)$
pairs, and in all cases its value is consistent with Eq.\ \eqref{x_BF}. The
order-parameter exponent $\beta$ can generally be evaluated to higher accuracy
than $\nu$ (because the latter depends on $T_l$ values obtained by interpolation
of data collected at a discrete set of temperature points), and has also been
calculated for almost all $(r,s)$ pairs. By contrast, we have explicitly
computed $1/\nu$ for only about half the pairs. In every case where it can be
tested, spanning the full range of bath exponents $0\le r<\half$ and
$\half\le s<1$, hyperscaling [Eq.\ \eqref{beta:hyper}] holds to within our
estimated numerical uncertainty. For this reason, we can confidently apply
Eq.\ \eqref{beta:hyper} to predict the value of $1/\nu$ in cases where it has
not been computed directly. Finally, we note that wherever it has been
calculated, $\delta$ [defined in Eq.\ \eqref{delta}, but not listed in Table
\ref{tab:NRG_exponents}] obeys Eq.\ \eqref{delta:hyper} to high precision.

Based on an examination of the critical spectra and the exponents $x$ and
$\beta$ (and hence $\nu$ via hyperscaling), we are able to identify three
distinct types of quantum criticality:

\noindent $\bullet$
Fermionic (F)---The asymptotic low-energy critical spectrum
exhibits SU(2) spin symmetry and decomposes into a direct
product of the spectrum of free bosons with bath exponent $s$ and the critical
spectrum of the pseudogap Kondo model with band exponent $r$, i.e., 
(BF critical) = (B free) $\otimes$ (F critical).
All static critical exponents that have been calculated are identical to
those of the pure-fermionic pseudogap Anderson and Kondo models with the same
$r$.

\noindent $\bullet$
Bosonic (B)---The asymptotic low-energy critical spectrum
exhibits SU(2) spin symmetry and decomposes into a direct
product of the critical spectrum of the spin-boson model with bath exponent $s$
and the strong-coupling spectrum\cite{phase_shift} of the pseudogap Kondo model
with band exponent $r$, i.e., (BF critical) = (B critical) $\otimes$
(F strong-coupling). All static static critical exponents that have been
calculated in region B are identical to those of the spin-boson model and of the
metallic ($r=0$) BFA and BFK models with the same $s$. However, despite the
asymptotic decomposition of the low-energy spectrum, the single-particle
spectral function has a non-Fermi liquid form (at least for $r=0$), as
discussed in Sec.\ \ref{subsec:dynamics}. This may be related to the fact that
the fermionic strong-coupling spectrum is not only found at the Kondo fixed
point where the exchange couplings take renormalized values
$J_z=J_{\perp}=\infty$, but also is approached for $J_z=\infty$, $J_{\perp}=$
finite in the limit of energy scales much smaller than $J_{\perp}$. We will
return to this observation in Sec.\ \ref{subsec:flows}.

\noindent $\bullet$
Mixed (M)---The critical spectrum exhibits broken SU(2) spin symmetry
and does not decompose into a direct product of
bosonic and fermionic parts. The exponents satisfy $x=x_B(s)=s$ but the
order-parameter exponent lies between the values for the spin-boson model
and the pseudogap Kondo/Anderson models, i.e.,
$\nu^{-1}_F(r)<\nu^{-1}(r,s)<\nu^{-1}_B(s)$.

\begin{figure}[t]
\includegraphics[height=2.2in]{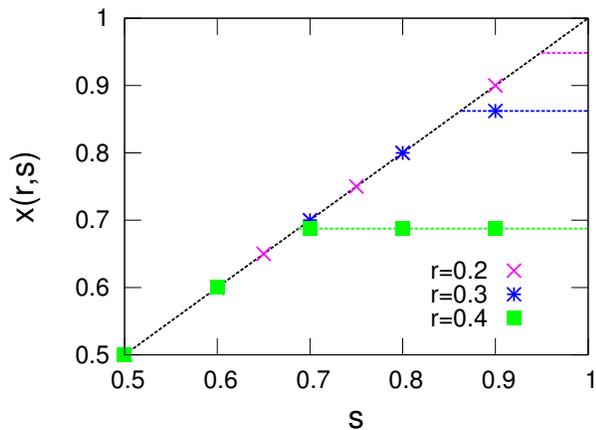}
\caption{\label{fig:x_vs_s} (Color online)
Static magnetic critical exponent $x$ vs bath exponent $s$ for $r=0.2$, $0.3$,
and $0.4$. The diagonal line represents the pure-bosonic exponent $x_B(s)=s$,
while each horizontal line segment shows the pure-fermionic value $x_F(r)$.
For each $(r,s)$ pair, the Bose-Fermi exponent satisfies
$x=\mathrm{min}[x_F(r),x_B(s)]$.}
\end{figure}

\begin{figure}[t]
\includegraphics[height=2.2in]{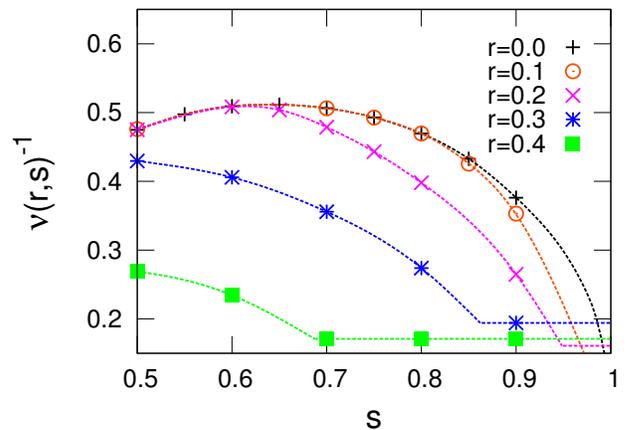}
\caption{\label{fig:nu_vs_s} (Color online)
Reciprocal of the correlation length exponent $1/\nu$ vs bath exponent $s$ for
the metallic case $r=0$ and for pseudogaps described by $r=0.1$--$0.4$.
The $r=0$ exponents coincide with those of the corresponding spin-boson
model and represent the values $1/\nu_B(s)$ describing pure-bosonic criticality.
Each horizontal line segment shows a pure-fermionic value $1/\nu_F(r)$. For each
$(r,s)$, the Bose-Fermi exponent coincides with $1/\nu_B(s)$ for $s<1-2r$
and with $1/\nu_F(r)$ for $s\ge x_F(r)$. For $1-2r<s<x_F(r)$, $1/\nu(r,s)$ lies
in between the bosonic and fermionic values.}
\end{figure}

The three types of quantum criticality are clearly revealed in plots of
$x$ and $1/\nu$ vs $s$ at fixed $r$. As can be seen in Fig.\ \ref{fig:x_vs_s},
$x(r,s)$ coincides with $x_B(s)=s$ until the latter value exceeds its
counterpart $x_F(r)$ in the pure-fermionic pseudogap problem. The break in the
slope of $x$ vs $s$ marks the transition from M-type to F-type criticality.
Fig.\ \ref{fig:nu_vs_s} shows that the Bose-Fermi correlation-length exponent
equals the bosonic value $\nu_B(s)$ for sufficiently small $s$ (B-type
criticality) and equals the fermionic value $\nu_F(r)$ over precisely the range
of $s$ where $x(r,s) = x_F(r)$ (F-type criticality), but between these regimes,
$\nu(r,s)$ takes values that differ from both $\nu_B(s)$ and $\nu_F(r)$
(M-type criticality).    

\begin{figure}[t]
\includegraphics[height=2.2in]{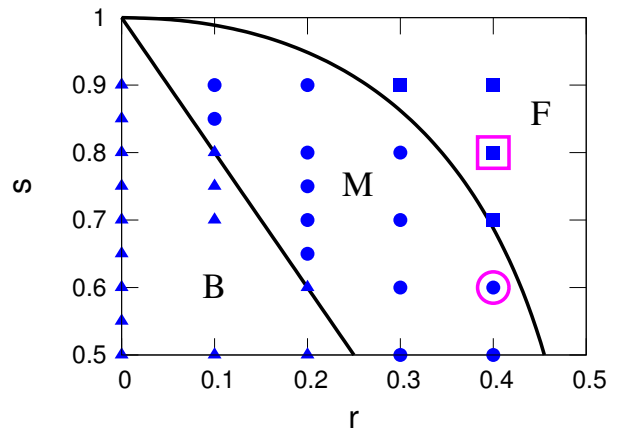}
\caption{\label{fig:r-s_plane} (Color online)
Summary of the pairs of bath exponents $(r,s)$ studied in this work. Results for
$r=0$ describe a metallic conduction band, while those for $r>0$ correspond to
pseudogapped problems. Squares, triangles, and circles respectively
correspond to quantum criticality of the F, B, and M types, as described in
the text. Filled symbols summarize NRG results for the BFK model while open
symbols represent CT-QMC results for the BFA model. Solid lines show the
conjectured boundaries $s=1-2r$ and $x_B(s)=s=x_F(r)$ between the different
types of criticality.}
\end{figure}

Figure \ref{fig:r-s_plane} summarizes the type of criticality found at different
locations on the $r$--$s$ plane, including points studied for the metallic
case $r=0$ where the behavior is always of the B type. It is seen that each
type of criticality (F, B, or M) occupies a contiguous region. All the results
are consistent with there being a boundary $s=x_F(r)$ between the F and M
regions (shown as a solid curve in Fig.\ \ref{fig:r-s_plane}). As argued at the
beginning of this section, such a boundary arises from the assumption that
the spin response at the Bose-Fermi critical point is dominated by the bath
(bosonic or fermionic) that has the more singular dynamical spin fluctuations,
corresponding to the smaller value of $x$.
The results in Fig.\ \ref{fig:r-s_plane} are also consistent with there being
a boundary $s=1-2r$ between the M and F regions (the straight line in the
figure). Such a boundary marks the line of equality of the frequency
exponents of the bare bosonic propagator and the fermionic particle-hole
bubble,\cite{Vojta:private} although the significance of this observation in
the present context remains to be established.

\subsection{Critical Dynamics}
\label{subsec:dynamics}

We now turn to the finite-temperature dynamics of the impurity Green's function
and the local susceptibility calculated at the critical bosonic coupling
$g=g_c$. Since the NRG is unreliable in the regime $|\omega|\lesssim T$, we
rely mainly on the CT-QMC for this part of the study, focusing once more on the
type-M case $r=0.4$, $s=0.6$ and on the type-F case $r=0.4$, $s=0.8$.

\begin{figure}[t]
\includegraphics[width=2.2in,angle=270]{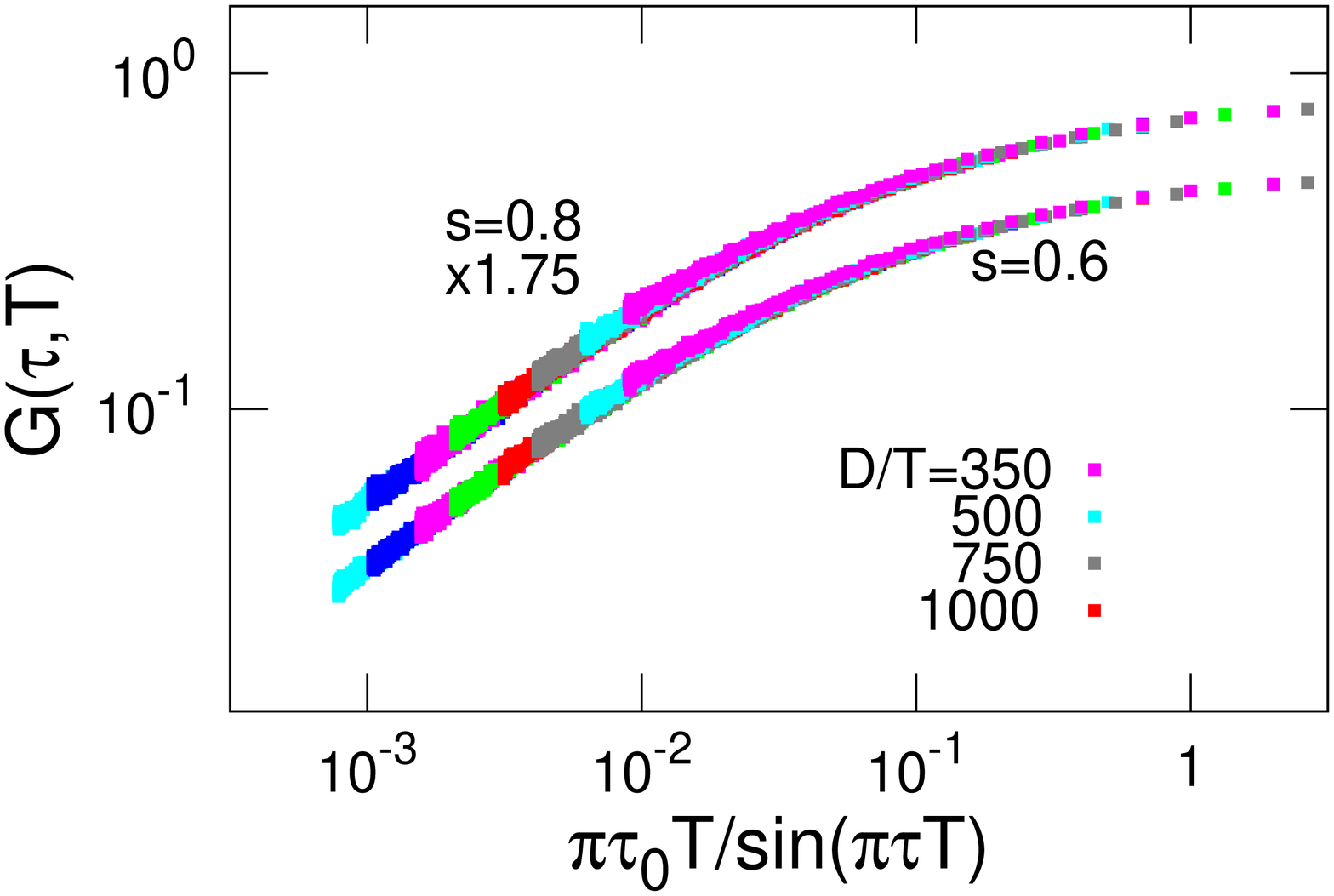}
\caption{\label{fig:G_vs_zeta} (Color online)
Single-particle Green's function $G(\tau,T)$ vs
$\xi=\pi \tau_0T/\sin(\pi\tau T)$ for $r=0.4$, $s=0.6$, $g=0.225D\approx g_c$
(lower data) and for $r=0.4$, $s=0.8$, $g=0.28D\approx g_c$ (upper data, all
$G$ values multiplied by 1.75 to avoid overlap with the $s=0.6$ data).
One observes excellent collapse of the data for just under two decades of
$\xi$. Temperature labels are shared between this figure and
Fig.\ \protect\ref{fig:chi_vs_zeta}.}
\end{figure}

\begin{figure}[t]
\includegraphics[width=2.2in,angle=270]{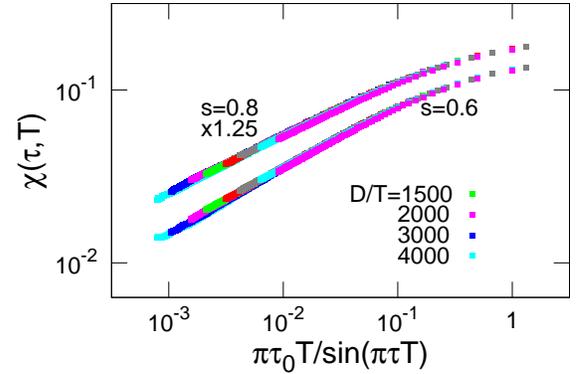}
\caption{\label{fig:chi_vs_zeta} (Color online)
Local spin susceptibility $\chi(\tau,T)$ vs $\xi=\pi\tau_0 T/\sin(\pi\tau T)$
for $r=0.4$, $s=0.6$, $g=0.225D\approx g_c$ (lower data) and for $r=0.4$,
$s=0.8$, $g=0.28D\approx g_c$ (upper data, all $\chi$ values multiplied by 1.25
to avoid overlap with the $s=0.6$ data). One observes excellent collapse of the
data for over two decades of
$\xi$. Temperature labels are shared between this figure and
Fig.\ \protect\ref{fig:G_vs_zeta}.}
\end{figure}

Figs.\ \ref{fig:G_vs_zeta} and \ref{fig:chi_vs_zeta} plot $G(\tau,T)$
and $\chi_{\text{loc}}(\tau,T)$, respectively, as functions of the combination
$\xi=\pi\tau_0 T/\sin(\pi\tau T)$, where $\tau_0=1/D$ renders the scaling
function dimensionless. For both $(r,s)$ pairs, the critical correlation
functions at temperatures well below the bare Kondo temperature,
$T_K^0\equiv T_K(g=0) \approx 0.06D$, exhibit excellent scaling collapse over
two decades of $\xi$. The scaling collapse leads to the important conclusions
that in the long-time limit $\tau T_K^0 \gg 1$,
\begin{align}
G(\tau,T) &=\Psi\biggl(\frac{\pi \tau_0T}{\sin(\pi \tau T)}\biggr)
  \stackrel{T\ll T_K^{0}}{\large \sim}\,
  \biggl(\frac{\pi\tau_0 T}{\sin(\pi \tau T)}\biggr)^{\eta_G} , \notag \\
\chi_{\loc}(\tau,T) &=\Phi\biggl(\frac{\pi \tau_0T}{\sin(\pi \tau T)}\biggr)
  \stackrel{T\ll T_K^{0}}{\large \sim}\,
  \biggl(\frac{\pi\tau_0 T}{\sin(\pi \tau T)}\biggr)^{\eta_{\chi}} .
\label{scalingformchi}
\end{align}
We obtain $\eta_G = 0.58(4)$, $\eta_{\chi}=0.40(2)$ for $s=0.6$ and
$\eta_G = 0.58(4)$, $\eta_{\chi}=0.31(2)$ for $s=0.8$.
These exponents are consistent with the relations
\begin{equation}
\label{etas}
\eta_G=1-r, \qquad
\eta_{\chi} = 1 - x,
\end{equation}
where $x(r,s)$ is defined in Eq.\ \eqref{chi_stat:crit}.
In the zero-temperature limit, Eq.\ \eqref{scalingformchi} and the first
Eq.\ \eqref{etas} give $G(\tau,T\rightarrow 0)\sim \tau^{-(1-r)}$,
reproducing an exact result.\cite{Vojta.03,Kircan.04} 
The quality of the scaling collapse in Figs.\ \ref{fig:G_vs_zeta} and
\ref{fig:chi_vs_zeta}, as well as the reproduction of the correct
zero-temperature limit, provide significant evidence that 
our results have reached the asymptotic low-energy scaling regime.

Since $1-r$ and $1-x$ are
both less then one, each correlator obeys $\omega/T$ scaling per the discussion
of Eq. \eqref{scaling_form}.
The observation that $\eta_{\chi}\ne2\eta_G$ implies that vertex corrections
cannot be neglected, in line with the fully interacting nature of the QCP.

For reasons discussed in Sec.\ \ref{sec:summary}, we have not been able to
access low enough temperatures using CT-QMC to study cases of B-type quantum
criticality. In this region of the $r$-$s$ plane, we must rely on information
from previous NRG studies of the $r=0$ BFK and BFA
models,\cite{Glossop.05,Glossop.07} which have demonstrated that at the
critical coupling, the local susceptibility calculated for
$|\omega|\gtrsim T$ is consistent with the existence of $\omega/T$ scaling,
while the zero-temperature impurity spectral function shows a non-Fermi liquid
form. The latter is also seen for $r=0$ results in a dynamical large-$N$
limit,\cite{Kirchner.05} as a function of both frequency and temperature.
It is reasonable to attribute this non-Fermi-liquid behavior to the
existence of an RG-irrelevant coupling between the fermionic and bosonic
sectors of the Hilbert space.

\subsection{RG Flow Diagrams}
\label{subsec:flows}

\begin{figure}[t]
\includegraphics[width=1.65in]{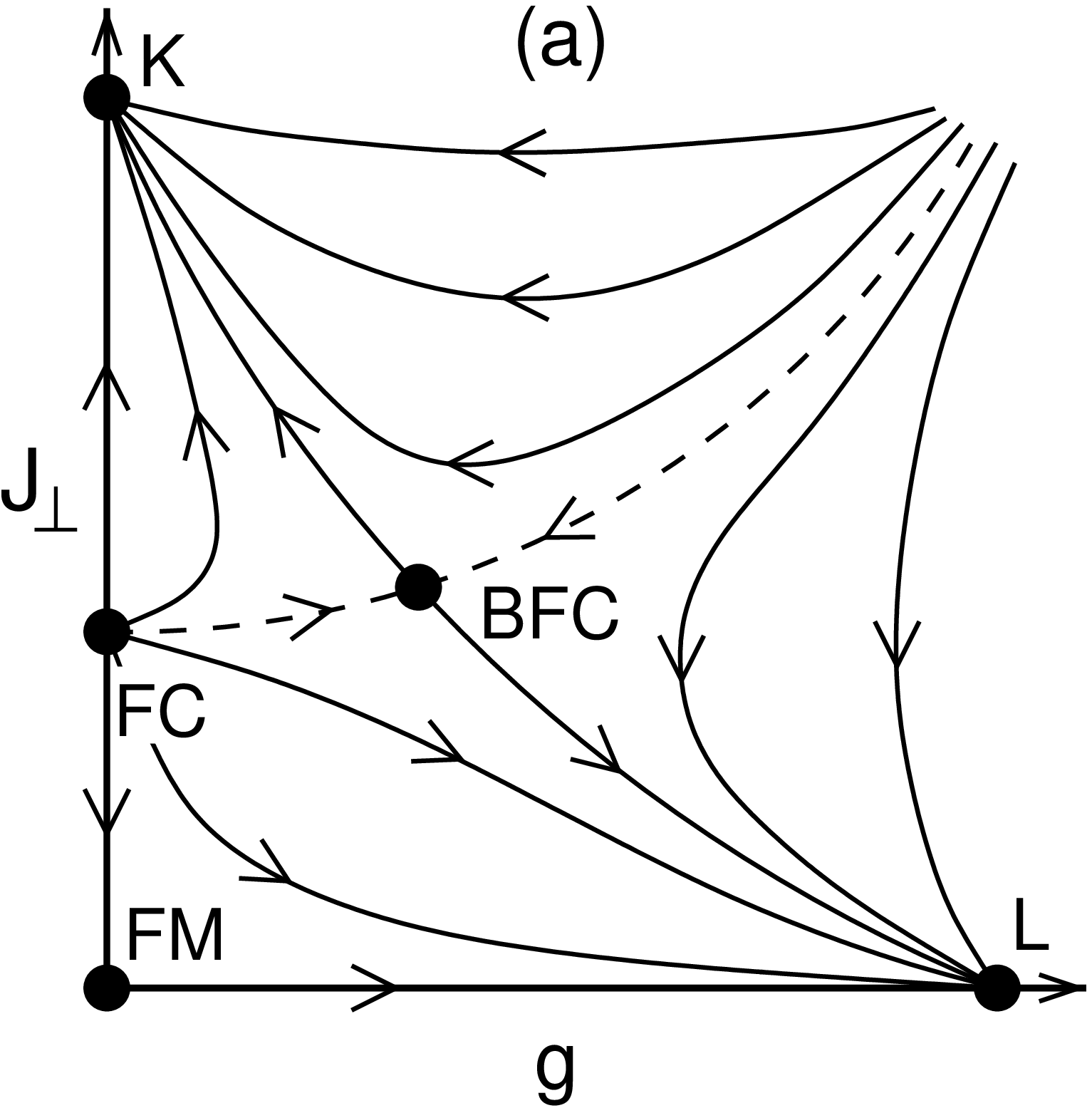}\!\!\!
\includegraphics[width=1.65in]{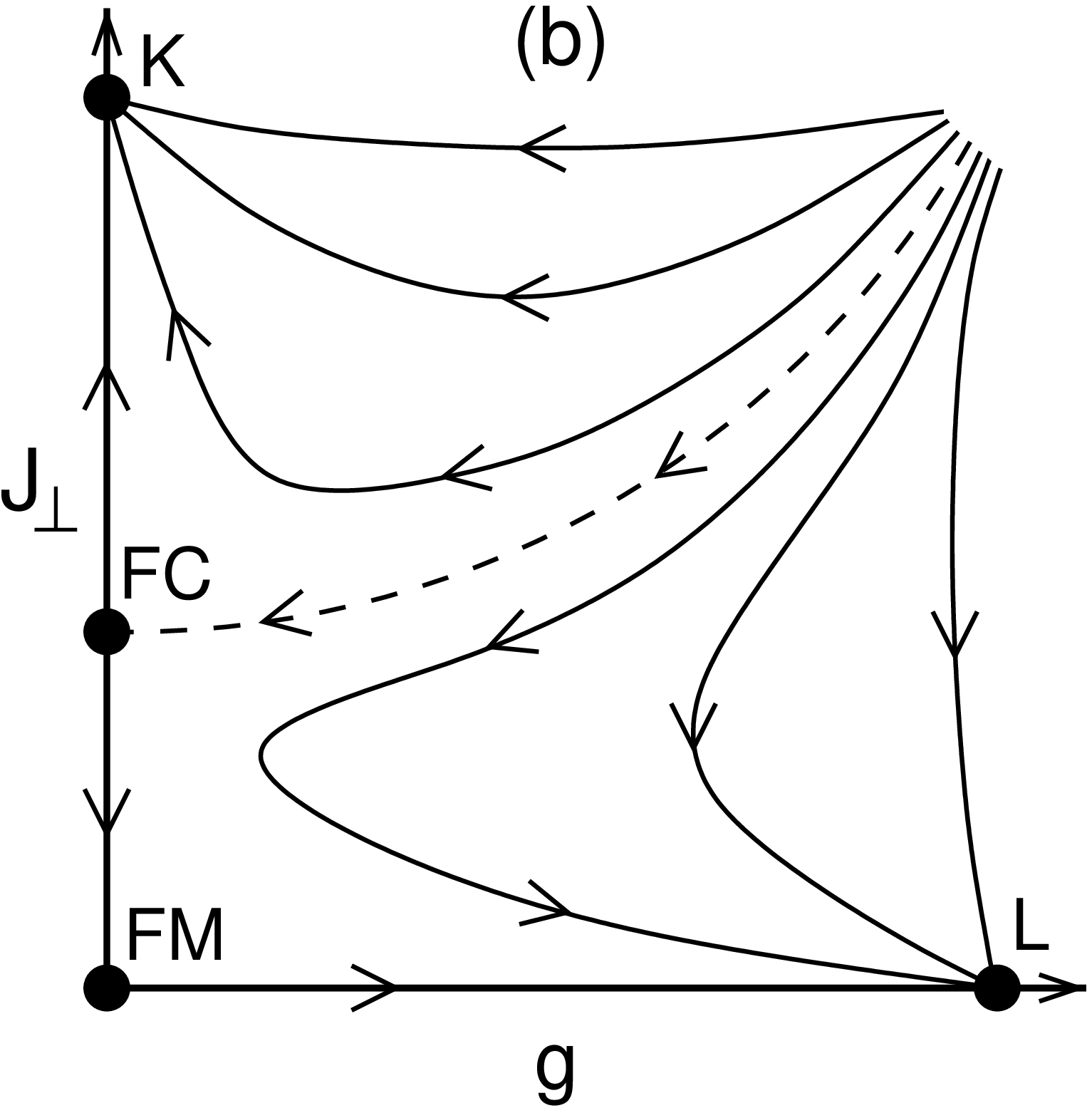}
\caption{\label{fig:RG-flows}
Schematic RG flow diagrams for the pseudogap BFK model projected onto the
plane spanned by the impurity-boson coupling $g$ and the spin-flip exchange
scattering $J_{\perp}$, valid for bath exponents $(r,s)$ such that the quantum
criticality is of the (a) B or M, and (b) F types. Not shown in these diagrams
is  a third axis describing the longitudinal exchange coupling $J_z$. The dashed
line marks the boundary between the Kondo and localized phases. Arrows show the
direction of RG flow of the effective couplings between fixed points represented
by circles: free moment (FM, at $J_z=0$), Kondo (K, at $J_z=\infty$), localized
(L, a line of fixed points spanning $0\le J_z\le\infty$), fermionic critical
(FC, at $J_z=J_{\perp}$), and Bose-Fermi critical (BFC, at $J_z=\infty$ in region
B but at finite $J_z>J_{\perp}$ in region M).
Very similar RG flow diagrams can be constructed for the pseudogap BFA model.}
\end{figure}

We conclude this section with a discussion of the RG structure of the
Ising-anisotropic pseudogap BFK model in the different regions of the $r$-$s$
plane. It is important to recognize that the bare couplings $(g,\,J)$ entering
Eq.\ \eqref{H_BFK} are insufficient to describe this RG structure. Since a
coupling $g>0$ breaks spin-rotation symmetry, SU(2) symmetry of the Kondo
exchange need not be preserved under renormalization, and a minimal description
involves keeping track of a triad of effective couplings
$(g,\,J_z,\,J_{\perp})$.

The stable fixed points of the pseudogap BFK model are the Kondo fixed point
at $g_K=0$, $J_{z,K}=J_{\perp,K}=\infty$ and a line of localized fixed points
at $g_L=\infty$, $J_{\perp,L}=0$, and (at least for $r=0$; see Ref.\
\onlinecite{Glossop.07}) $J_{z,L}\propto (g - g_c)^{-\beta}$ where $g$ is the
bare bosonic coupling and $\beta$ is the order-parameter exponent.
The QCP of the pseudogap Kondo model is located at $g_F=0$ and
$J_{z,F} = J_{\perp,F} = J_c(g=0)$, where $\rho_0 J_c \to r$ for $r\to 0$.

A perturbative RG treatment of the model valid for $r=0$ and $0<1-s\ll 1$
has shown\cite{Zhu.02} that the Bose-Fermi QCP is located at $K_0 g^*=O(1)$,
$J_z^* = \infty$, and $\rho_0 J_{\perp}^*\simeq \sqrt{1-s} \ll 1$, with the
NRG having confirmed the result $J_z^*=\infty$ over a wider range of $s$
values.\cite{Glossop.07} What remains to be established is the manner in which
$(g^*,\,J_z^*,\,J_{\perp}^*)$ evolves with increasing $r$ and/or $s$ to reach
$(0,J_c(0),J_c(0))$ upon entry into the region F of the $r$-$s$ plane. Since
region F is bounded by the line $x_F(r)=s$, and the exponent $x_F(r)$ is very
well described\cite{Ingersent.02} by
$x_F=1-[\rho_0 J_c(0)]^2=1-[\rho_0 J_{\perp}^*]^2$, one is led
to conclude that for a given value of $s$ satisfying $0<1-s\ll 1$, the relation
$\rho_0 J_{\perp}^* \simeq\sqrt{1-s}$ holds true both at $r=0$ (on the left-most
edge of region B in Fig.\ \ref{fig:r-s_plane}) and at the border between regions
M and F. This observation, when combined with the asympototic spectral
decomposition (BF critical) = (B critical) $\otimes$ (F strong-coupling) that
holds throughout region B, naturally leads to the conjecture that upon
increasing $r$ from 0 to the point of entry into region F, (1) $J_{\perp}^*$
remains constant (or very nearly so); (2) $K_0 g^*$ decreases monotonically
from a value of order unity to reach zero; and (3) $J_z^*$ is infinite
throughout region B, and decreases monotonically to $J_c(r)$ on crossing
region M. The presence in region M of finite exchange couplings satisfying
$J_z^*>J_{\perp}^*$ is consistent with the observation of broken SU(2) spin
symmetry in the asymptotic low-energy critical spectrum. By contrast, the
value $J_z^*=\infty$ in region B ensures that the spectrum appears to be
SU(2) invariant at energy scales much below $J_{\perp}^*$.

Support for this picture comes from Table \ref{tab:NRG_exponents}, which
shows a clear trend with increasing $r$ at fixed $s$ (or with increasing $s$ at
fixed $r$) in the value $g_u$ of the bosonic coupling that brings the model
into its quantum-critical regime at the highest temperature. Throughout region
B in Fig.\ \ref{fig:r-s_plane}, $g_u$ and the corresponding Kondo coupling
$J_u = J_c(g_u)$ are very large in order to achieve rapid flow to $J_z^*=\infty$
and the large fixed-point value of $K_0 g^*$. While region M is crossed, $g_u$
and $J_u$ decrease in line with $g^*$ and $J_z^*$. In region F, the value
$g_u=0$ shows that the critical point can be reached without any coupling of
the impurity to the bosons, meaning that $g^*$ is necessarily zero and that
$J_z^*=J_{\perp}^*=J_u$.

Based on the preceding considerations we propose the schematic RG flow
diagrams shown projected onto the $g$-$J_{\perp}$ plane in Fig.\
\ref{fig:RG-flows}. In regions B and M, a QCP labeled BFC in Fig.\
\ref{fig:RG-flows}(a) lies on the separatrix between the basins of attraction
of the Kondo and localized fixed points (K and L, respectively). RG flow along
the separatrix is toward BFC, and on the small-$g$ side, away from the fermionic
pseudogap critical point FC. With increasing $r$ at fixed $s$ (or increasing $s$
at fixed $r$), BFC moves to smaller values of $g$, and merges with FC at
the boundary between regions M and F.
Throughout the latter region of the $r$-$s$ plane, the RG structure is as shown
in Fig.\ \ref{fig:RG-flows}(b), with flow along the separatrix toward FC.

\subsection{Effect of Particle-Hole Asymmetry}
\label{subsec:p-h-asymm}

To this point, we have focused exclusively on conditions of strict
particle-hole (\ph) symmetry, i.e., $U=-2\epsilon_d$ for the BFA model,
which maps to a BFK model with potential scattering $W=0$. In this section,
we consider the effects of breaking this symmmetry, supporting our arguments
with NRG results for the critical spectrum and for the critical exponent $x$
entering Eq.\ \eqref{chi_stat:crit}, as obtained in a preliminary study of the
BFK model with $W\ne 0$.

Under the conditions of \ph\ symmetry defined in the previous paragraph,
the pseudogap Kondo and Anderson models exhibit a Kondo-destruction QCP only
over the range of band exponents $0<r<\half$; for $r\ge\half$, the Kondo phase
disappears and over the entire parameter space the system approaches the
FM fixed point at low temperatures.
In these models, it is known that away from \ph\ symmetry, a
strong-coupling or Kondo phase is present for all $r>0$, and that for $0<r<1$
this phase is separated from the free-moment phase by an interacting QCP.
For exponents $0<r<r^*\approx0.375$, \ph\ asymmetry is irrelevant at the QCP
and the quantum criticality is identical to that at \ph\ symmetry, whereas for
$r^* < r < 1$, quantum criticality away from \ph\ symmetry is governed by a
distinct asymmetric QCP.\cite{Buxton.98}

Combining these well-established properties of the pure-fermionic
pseudogap models with the results presented in Sec.\ \ref{sec:critical}
allows informed speculation about the effects of \ph\ asymmetry in the
pseudogap BFA and BFK models. We expect the persistence of a region of
B type criticality in which the critical spectrum has the product form
(BF critical) = (B critical) $\otimes$ (F strong-coupling), with \ph\ asymmetry
affecting the fermionic strong-coupling spectrum but not the critical
properties studied in this work, all of which are determined solely by
the bosonic spectrum. For $r=0$, \ph\ asymmetry should be marginal (as
it is in the metallic Anderson and Kondo models), giving rise to a line of
QCPs sharing the same critical properties. For $r>0$, by contrast,
\ph\ asymmetry should be relevant, with the fermionic spectrum being
that of the asymmetric strong-coupling fixed point. These conjectures
are consistent with preliminary NRG studies of the cases $(r,s)=(0,0.7)$
and $(0.1,0.7)$.

In the presence of \ph\ asymmetry, we also expect a region in which the
dynamical spin response arising from the fermions is more singular than that
from the bosonic bath. This region of what we will dub F$^{\prime}$-type
criticality is likely to span the range of exponents $0<r<1$ and $\half<r<1$
in which $s>x_F'(r)$. Here, $x_F'(r)$ is the value of the exponent $x$ in
Eq.\ \eqref{chi_stat:crit} at the asymmetric pseudogap QCP. For $0<r<r^*$,
$x_F'=x_F$ (Ref.\ \onlinecite{Ingersent.02}) so F$^{\prime}$ criticality
should be identical to F-type. For $r^*<r<1$, by contrast, $x_F'>x_F$ (Ref.\
\onlinecite{Ingersent.02}), and the F$^{\prime}$ critical exponents
should belong to the universality class of the asymmetric pseudogap QCP.
We have found an example of F$^{\prime}$ criticality for $(r,s)=(0.4,0.9)$.

Finally, it seems probable that the $B$ and $F^{\prime}$ regions will
be separated by one in which the critical spectrum does not have a simple
direct product form. At \ph\ symmetry, the M-type region covers the range
$1-2r<s<x_F(r)$. As discussed in Sec.\ \ref{subsubsec:other-r,s}, the lower
bound on the range of $s$ seems to be defined by the equality of the
frequency exponents of the bare bosonic propgator and the fermionic
particle-hole bubble. Since \ph\ asymmetry is irrelevant at the FM fixed
point,\cite{Buxton.98} there seems to be no reason to expect the boundary to
be affected by this breaking of symmetry. On the other hand, the upper bound
$s=x_F(r)$ seems likely to be replaced by $s=x_F'(r)$, the condition discussed
in the previous paragraph for entry into a region of F$^{\prime}$ criticality.

\section{Summary}
\label{sec:summary}

In this work, we have applied a combination of continuous-time quantum
Monte-Carlo (CT-QMC) and numerical renormalization-group (NRG) methods to study
systematically the interplay between bosonic and fermionic baths, each of which
on its own can induce critical Kondo destruction at a continuous
zero-temperature transition. We have shown that at particle-hole symmetry, the
quantum critical point (QCP) in the easy-axis pseudogap BFA model belongs to
the same universality class as the QCP of the corresponding Kondo model. 
We have further shown the surprising result that the value of the exponent $x$
for the temperature dependence of the critical local spin susceptibility
[defined in Eq.\ \eqref{chi_stat:crit}] of either model is sensitive to the
exponents $r$ and $s$ characterizing the vanishing of the fermionic and bosonic
densities of states.  In the region of the $r$-$s$ plane where
$s\ge x_F(r)$ ($x_F$ being the thermal critical
exponent of the pseudogap Anderson and Kondo models without bosons),
all critical exponents of the Bose-Fermi models that we have calculated
are identical to those of the pure-fermionic models, and the critical many-body
spectrum decomposes into a direct product of a free bosonic spectrum and a
critical pseudogap fermionic spectrum; this regime has eluded all previous
studies. For $s<x_F(r)$, the critical spin fluctuations are instead
dominated by the bosonic bath, leading to $x=x_B(s)=s$. However, the
correlation-length exponent $\nu$ [defined in Eq. \eqref{nu}] coincides with
that of the spin-boson model, and the asymptotic low-energy critical spectrum
decomposes into a direct product of a critical bosonic spectrum and a pseudogap
Kondo fermionic spectrum, only for $s\le 1-2r$. Within an intermediate region
$1-2r<s<x_F(r)$, $\nu(r,s)$ takes a value lying between those for the spin-boson
model and for the pseudogap Anderson and Kondo models, and bosonic and
fermionic degrees of freedom cannot be disentangled in the critical spectrum.
In all three regions, other static critical exponents are related to $x$ and
$\nu$ via hyperscaling relations that are expected to hold only at an
interacting critical point.

We have also shown that at the QCP, the imaginary-time correlation functions
$G(\tau, T)$ and $\chi_{\loc}(\tau, T)$ scale as functions of
$\xi=\pi T \tau_0/\sin(\pi\tau T)$ and that their real-frequency counterparts
obey $\omega/T$ scaling, consistent with the notion that the QCP is fully
interacting. Scaling collapse of imaginary-time correlators as functions of
$\xi$ has previously been reported for the sub-Ohmic BFK
model,\cite{Kirchner.08} where the Kondo effect is critically destroyed by the
bosonic bath, and in the pseudogap Anderson model at and away from
particle-hole symmetry,\cite{Glossop.11,Pixley.11} where criticality is driven
by fermionic fluctuations of the band. That it generalizes to the more complex
case considered in the present work suggests that the scaling collapse may very
well be a general feature of local quantum criticality. The scaling collapse in
terms of $\xi$ implies (under conditions spelled out in Sec.\
\ref{subsec:dynamics}) that the associated real-frequency correlator displays
$\omega/T$ scaling, a property that has been reported in several experiments on
unconventional quantum criticality in $4f$-electron based
magnets.\cite{Schroeder.00,Friedemann.09}
A scaling collapse of the form observed here is natural for boundary-conformal
quantum impurity systems. However, conformal symmetry is broken both by the
bosonic bath (as in the sub-Ohmic BFK model), and the fermionic bath (as in
the pseudogap Anderson model).
Symmetry restoration frequently accompanies criticality. The case discussed
here, however, has to be distinguished from this more standard situation of
irrelevant symmetry-breaking fields, as the broken symmetry in the bulk induces
boundary criticality with local correlators that are compatible with a boundary
conformal critical theory. As discussed earlier,\cite{Kirchner.08} a deeper
understanding of this observation should help identify a critical field theory
of unconventional quantum criticality.

From a methodological viewpoint, this work has shown that the CT-QMC method can
attain sufficiently low temperatures in the presence of a bosonic bath to
identify a quantum critical point lying between stable phases, and to obtain
critical properties in agreement with those given by the NRG. Such study is
possible using the CT-QMC only in cases where entry into the critical power-law
regime takes place at a fairly high temperature $T_u$. The presence of a
pseudogap in the fermionic density of states helps in this regard: the value
of $T_u$ decreases as $r$ decreases, so larger $r$ values are optimal.
Decreasing the bath exponent $s$ also reduces $T_u$, making it difficult to
study the range $s<\half$ without fine-tuning of the Hamiltonian.

\acknowledgments

We thank M.\ T.\ Glossop, P.\ Ribeiro, M.\ Vojta, F.\ Zamani, and J.-X.\ Zhu for
useful discussions. This work has been supported by NSF Grants No.\ DMR-0710540,
DMR-1309531, and DMR-1107814, the Robert A.\ Welch Foundation Grant No.\ C-1411,
and the Alexander von Humboldt Foundation. We acknowledge the hospitality of
the Max Planck Institute for the Physics of Complex Systems (K.I.\ and J.H.P.),
the Aspen Center for Physics (NSF Grant No.\ 1066293; S.K.\ and Q.S.), the
Institute of Physics of Chinese Academy of Sciences, and Karlsruhe Institute
of Technology (Q.S.).


\begin{thebibliography}{30}
\bibitem{units}
Throughout this paper, we adopt units where $\hbar = k_B = 1$.
\bibitem[\citenamefont{note1}()]{note1}
\bibinfo{note} For $2\lambda\geq 1$, the Fourier-transform integral does not
converge, necessitating introduction of a finite cutoff scale that spoils
$\omega/T$ scaling.
\bibitem[\citenamefont{multiplets}()]{multiplets}
\bibinfo{note}
In problems that exhibit strict particle-hole symmetry, the U(1) symmetry
that leads to conservation of total charge (electron number measured from
half-filling) $Q_{\tot} = \sum_{n,\sigma} (f_{n\sigma}^{\dag} f_{n\sigma} -
\half)$ is promoted to an SU(2) isospin symmetry generated by
$I_{\tot}^z \equiv \half Q_{\tot}$ and $I_{\tot}^+ \equiv (I_{\tot}^-)^{\dag}
= \sum_n (-1)^n f_{n\uparrow}^{\dag} f_{n\downarrow}^{\dag}$.
The NRG calculations were performed on a basis of multiplets labeled by quantum
numbers $S_{\tot}^z$ (the $z$ component of the total spin) and $I_{\tot}$,
with each multiplet representing $2 I_{\tot}+1$ degenerate eigenstates.
\bibitem{phase_shift}
The strong-coupling spectrum of the particle-hole-symmetric pseudogap
Anderson and Kondo models is characterized by a phase shift of $(1-r)\pi/2$
for electrons at the Fermi energy; see Ref.\ \onlinecite{Buxton.98}.
\end{thebibliography}
\end{document}